\documentclass{article}

\usepackage{titling}
\usepackage[centertags]{amsmath}
\usepackage{chicago,fancyhdr}
\usepackage{amssymb,amsfonts,eucal,oldgerm}
\usepackage[english]{babel}
\usepackage[text={6in,8.25in},centering]{geometry}

\usepackage[colorlinks=true,citecolor=red,linkcolor=blue,urlcolor=blue,pdftex]{hyperref}

\newtheorem{theorem}{Theorem}[subsection]

\newtheorem{conjecture}[theorem]{Conjecture}

\newcommand{\skipline}[1][1]{\vspace*{#1\baselineskip}}
\newcommand{\half}{\frac{1}{2}}
\newcommand{\athird}{\frac{1}{3}}

\newcommand{\resetcounters}{%
  \setcounter{equation}{0}%
  \setcounter{theorem}{0}%
  \setcounter{figure}{0}%
}
\newcommand{\resetsec}{%
  \resetcounters%
  \renewcommand{\thetheorem}{\arabic{section}.\arabic{theorem}}%
  \renewcommand{\theequation}{\arabic{section}.\arabic{equation}}%
}
\newcommand{\resetsub}{%
  \resetcounters%
  \renewcommand{\thetheorem}{\arabic{section}.\arabic{subsection}%
    .\arabic{theorem}}%
  \renewcommand{\theequation}{\arabic{section}.\arabic{subsection}%
    .\arabic{equation}}%
  \renewcommand{\thefigure}{\arabic{section}.\arabic{subsection}%
    .\arabic{figure}}%
}

\newcommand{\coloneq}{\mathrel{\mathop:}=}
\newcommand{\nth}[1]{#1$^{\text{th}}$}

\title{A Primer on Energy Conditions\thanks{I thank Harvey Brown and
    David Malament for enjoyable discussions that first put me on the
    path to a thorough investigation of energy conditions.  I thank
    The Young Guns of the Spacetime Church of the Angle Brackets for
    suffering through a much longer version of this paper and giving
    me insightful help with a smile.}}

\author{Erik Curiel\thanks{\textbf{Author's address}: Munich Center
    for Mathematical Philosophy, Ludwig-Maximilians-Universit\"at,
    Ludwigstra{\ss}e 31, 80539 M\"unchen, Germany; \textbf{email}:
    \texttt{erik@strangebeautiful.com}}}

\begin{document}

\maketitle

\skipline

\begin{quote}
  \begin{center}
    \textbf{ABSTRACT}
  \end{center}  

  An energy condition, in the context of a wide class of spacetime
  theories (including general relativity), is, crudely speaking, a
  relation one demands the stress-energy tensor of matter satisfy in
  order to try to capture the idea that ``energy should be positive''.
  The remarkable fact I will discuss in this paper is that such
  simple, general, almost trivial seeming propositions have profound
  and far-reaching import for our understanding of the structure of
  relativistic spacetimes.  It is therefore especially surprising when
  one also learns that we have no clear understanding of the nature of
  these conditions, what theoretical status they have with respect to
  fundamental physics, what epistemic status they may have, when we
  should and should not expect them to be satisfied, and even in many
  cases how they and their consequences should be interpreted
  physically.  Or so I shall argue, by a detailed analysis of the
  technical and conceptual character of all the standard conditions
  used in physics today, including examination of their consequences
  and the circumstances in which they are believed to be violated.
\end{quote}

\newpage

\thispagestyle{empty}
\tableofcontents

\skipline

\section{The Character of Energy Conditions}
\label{sec:character}

An energy condition, in the context of a wide class of spacetime
theories (including general relativity),\footnote{From hereon until
  \S\ref{sec:constraints-char-st-theors}, unless explicitly stated
  otherwise, the discussion should be understood to be restricted to
  the context of general relativity.  Almost everything I say until
  then will in fact hold in a very wide class of spacetime theories,
  but the fixed context will greatly simplify the exposition.  In
  general relativity, the fundamental theoretical unit, so to speak,
  is a spacetime model consisting of an ordered pair $(\mathcal{M}, \,
  g_{ab})$, where $\mathcal{M}$ is a four-dimensional, paracompact,
  Hausdorff, connected, differential manifold and $g_{ab}$ is a
  pseudo-Riemannian metric on it of Lorentzian signature.  `$T_{ab}$'
  will always refer to the stress-energy tensor picked out in a
  spacetime model by the Einstein field equation, `$T$' to the trace
  of $T_{ab}$ ($T^n {}_n$), `$R_{ab}$' to the Ricci tensor associated
  with the Riemann tensor $R^a {}_{bcd}$ associated with the unique
  torsion-free derivative operator $\nabla$ associated with $g_{ab}$,
  `$R$' to the trace of the Ricci tensor ($R^n {}_n$, the Gaussian
  scalar curvature), and `$G_{ab}$' to the Einstein tensor ($R_{ab} -
  \half R g_{ab}$).  For conventions about the metric signature and
  the exact definitions of these tensors, I follow
  \citeN{malament-fnds-gr-ngt}.  Unless otherwise explicitly noted,
  indicial lower-case Latin letters ($a, \, b, \ldots$) designate
  abstract tensor-indices, indicial lower-case Greek letters ($\mu, \,
  \nu, \ldots$) designate components with respect to a fixed
  coordinate system or tetrad of tangent vectors ($\mu \in \{0, \, 1,
  \, 2, \, 3\}$), and hatted indicial lower-case Greek letters
  ($\hat{\mu}, \, \hat{\nu}, \ldots$) designate the spacelike
  components ($\hat{\mu} \in \{1, \, 2, \, 3\}$) with respect to a
  fixed $1 + 3$ tetrad system.  (For an exposition of the
  abstract-index notation, see \citeNP{penrose-rindler-spinors-st-1},
  \citeNP{wald-gr}, or \citeNP{malament-fnds-gr-ngt}.)}  is, crudely
speaking, a relation one demands the stress-energy tensor of matter
satisfy in order to try to capture the idea that ``energy should be
positive''.\footnote{There does not exist in general relativity a
  satisfactory definition for a ``gravitational'' stress-energy
  tensor, one that represents localized stress-energy of purely
  ``gravitational'' systems.  (See \citeNP{curiel-no-sab-phys}.)  One
  may want to think of this as a limitation on the possible physical
  content of the standard pointilliste energy conditions, as I discuss
  at the end of \S\ref{sec:point-energy-cond}.}  Perhaps the simplest
example is the so-called weak energy condition: for any timelike
vector $\xi^a$ at any point of the spacetime manifold, the
stress-energy tensor $T_{ab}$ satisfies $T_{mn} \xi^m \xi^n \geq 0$.
This has a simple physical interpretation: the (ordinary) energy
density of the fields contributing to $T_{ab}$, as measured in a
natural way by any observer (\emph{e}.\emph{g}., using instruments at
rest relative to that observer), is never negative.  The remarkable
fact I will discuss in this paper is that such simple, general, almost
trivial seeming propositions have profound and far-reaching import for
our understanding of the structure of relativistic spacetimes.  It is
therefore especially surprising when one also learns that we have no
clear understanding of the nature of these conditions, what
theoretical status they have \emph{vis}-\emph{\`a}-\emph{vis}
fundamental physics, what epistemic status they may have, when we
should and should not expect them to be satisfied, and even in many
cases how they and their consequences should be interpreted
physically.  Or so I shall argue.

\citeN[p.~260]{geroch-horowitz-glob-struc}, in discussing the form of
singularity theorems in general relativity, outline perhaps the most
fundamental reason for the importance of energy conditions with the
following pregnant observation:\label{pg:geroch-horo}
\begin{quote}
  One would of course have to impose \emph{some} restriction on the
  stress-energy of matter in order to obtain any singularity theorems,
  for with no restrictions Einstein's equation has no content.  One
  might have thought, however, that only a detailed specification of
  the stress-energy at each point would suffice,
  \emph{e}.\emph{g}.\ that one might have to prove a separate theorem
  for each combination of the innumerable substances which could be
  introduced into spacetime.  It is the energy condition which
  intervenes to make this subject simple.  On the one hand it seems to
  be a physically reasonable condition on all types of classical
  matter, while on the other it is precisely the condition on the
  matter one needs for the singularity theorem.
\end{quote}
I will return to this quote later, in
\S\ref{sec:constraints-char-st-theors}, but for now the salient point
is that a generic condition one imposes on the stress-energy tensor,
``generic'' in the sense that it can be formulated independently of
the details of the internal structure of the tensor, which is to say
independently of any quantitative or structural feature or
idiosyncrasy of any particular matter fields, suffices to prove
theorems of great depth and scope.  Indeed, as Geroch and Horowitz
suggest, without the possibility of relying on conditions of such a
generic character, we would not have the extraordinarily general and
far-reaching singularity theorems we do have.  And it is not only
singularity theorems that rely for their scope and power on these
energy conditions---it is no exaggeration to say that the great
renaissance in the study of general relativity itself that started in
the 1950s with the work of Synge, Wheeler, Misner, Sachs, Bondi,
Pirani, \emph{et al}., and the blossoming of the investigation of the
global structure of relativistic spacetimes at the hands of Penrose,
Hawking, Geroch, \emph{et al}., could not have happened without the
formulation and use of such energy conditions.

What is perhaps even more remarkable is that many of the most profound
results in the study of global structure---\emph{e}.\emph{g}., the
Hawking Area Theorem---do not depend on the Einstein field equation at
all, but rather assume only a purely formal condition imposed on the
Ricci tensor, which itself can be thought of as an ``energy''
condition if one invokes the Einstein field equation to provide a
physical interpretation of the Ricci tensor.  In a sense, therefore,
energy conditions seem to reach down to and get a hold of a level of
structure in our understanding of gravitation and relativistic
spacetimes even more fundamental than the Einstein field equation
itself.  (I will discuss in \S\ref{sec:constraints-char-st-theors}
this idea of ``levels of structure'' in our understanding of general
relativity in particular, and of gravitation and spacetime more
generally.)

Now, most propositions of a fundamental character in general
relativity admit of interpretation as either a postulate of the theory
or as a derived consequence from some other propositions taken as
postulates.  That is to say, the theory allows one a great deal of
freedom in what one will take as given and what one will demand a
proof of.  One can, for example, either assume the so-called Geodesic
Principle from the start as a fundamental regulative principle of the
theory, as, for example, in the exposition of
\citeN{malament-fnds-gr-ngt}, or one can assume other propositions as
fundamental, perhaps ones fixing the behavior of ideal clocks and
rods, and derive the Geodesic Principle as a consequence of those
propositions, as, for example, in the exposition of
\citeN{eddington-math-theor-rel}.  Which way one goes for any given
proposition depends, in general, on the context one is working in, the
aims of one's investigation, one's physical and philosophical
intuitions and predilections, \emph{etc}.\footnote{See \citeN[this
  volume]{weatherall-inert-mot-expl-fnds-st-thrs} for an insightful
  discussion of a view of the foundations of spacetime theories I find
  sympathetic and amenable to my own views as I sketch them here.}

This interpretive flexibility does not seem to hold, however, for
energy conditions.  I know of no substantive proposition that,
starting from some set of other important ``fundamental postulates'',
has as its consequence an energy condition.  One either imposes an
energy condition by fiat, or one shows that it holds for stress-energy
tensors associated with particular forms of matter fields.  One never
imposes general conditions on other geometrical structures
(\emph{e}.\emph{g}., the Riemann tensor or the topology or the global
causal structure) and derives therefrom the satisfaction of an energy
condition (except in the trivial case where one imposes conditions
directly on the Ricci or Einstein tensor, standing as a direct proxy
for the stress-energy tensor by dint of the relation between them
embodied by the Einstein field-equation).\footnote{The one possible
  exception to this claim I know of is the attempt by
  \citeN{wall-aanec-genl-2nd-law} to derive the so-called averaged
  null energy condition (ANEC) from the Generalized Second Law of
  thermodynamics.  While I find his arguments of great interest, I
  also find them problematic at best.  See
  \citeN{curiel-econds-qftcst} for discussion.}  There are a plethora
of results that show when various energy conditions may or must be
violated both theoretically and according to observation, which I
discuss in \S\ref{sec:violations}, but none that show non-trivially
when one must hold.  Indeed, this inability to prove them is an
essential part of what seems to make them structure ``at a deeper
level'' perhaps even than causality conditions (many of which can be
derived from other fundamental assumptions), and so applicable across
a \emph{very} wide range of possible theories of spacetime.

In a similar vein, they occupy an odd methodological and theoretical
niche quite generally.  None is implied by any known general theory,
though each can be formulated in the frameworks of a wide spectrum of
different theories, and several can be shown to be inconsistent with a
wide spectrum of theories (in the strong sense that one can derive
their respective negations in the context of the theories).  Indeed,
they are among the very few physical propositions I know that can be
used either to exclude as physically unreasonable individual solutions
to the field equations of a particular theory (as for,
\emph{e}.\emph{g}., a wide class of FLRW spacetimes in general
relativity that have strongly negative pressures\footnote{See
  \citeN{curiel-econds-cosmo} for discussion.}), or to exclude entire
theories (such as the Hoyle-Bondi steady-state theory of cosmology, as
I discuss below in \S\ref{sec:violations}).  Whether or not one should
consider them as ``part of'' any given theory, therefore, seems a
problematic question at best, and an ill-posed one at worst.

It is difficult to get a grasp on their epistemic status as well.
They seem in no sense to be laws, under any standard account in the
literature, for none of them holds for all known ``physically
reasonable'' types of matter, and each of them is in fact violated in
what seem to be physically important circumstances.  Neither do they
appear to be empirical or inductive generalizations, for the same
reason.\footnote{It should be noted, however, that, to the best of my
  knowledge, there has never been direct experimental observation of a
  violation of any of the standard energy conditions I discuss in
  \S\ref{sec:standard-energy-conds}.  We do, however, have extremely
  good \emph{indirect} experimental and observational evidence for
  violations of several of them, as I will discuss in
  \S\ref{sec:conseq-viol}.  See \citeN{curiel-econds-cosmo} for an
  extended discussion of evidence for their violation in cosmology.
  Even direct experimental verification of the Casimir effect does not
  yield direct measurement of negative energy densities, though the
  Casimir effect relies essentially on the existence of such; rather,
  the negative energy densities are inferred from measurement of the
  Casimir force itself
  \cite{brown-maclay-vac-stress-conduct-plates-img-soln}.}  And yet we
think that (at least) one of them---or something close to
them---likely holds generically in the actual universe, at the level
of classical (\emph{i}.\emph{e}., non-quantum) physics at least, and
even that one or more of them, appropriately reformulated, should hold
generically at the quantum level as well.\footnote{See
  \citeN{curiel-econds-qftcst} for discussion.}  Even more, as I have
already indicated, there seem to be very good reasons for thinking
that the sense in which they do obtain, whatever that may be, is
grounded in structure at a level of our understanding even deeper than
the Einstein field equation itself, which we surely do think of as a
law, under any reasonable construal of the notion.

So what are they?  The remainder of this paper consists of an attempt
to come to grips with this question, by exploring their formulations,
their consequences, their relations to other fundamental structures
and principles, and their role in constraining the possible forms a
viable theory of spacetime may take.  Those who hope for a decisive
answer to the question will leave disappointed.  I feel I will have
succeeded well enough if I am able only to survey the most important
issues and questions, clarify and sharpen some of them, propose a few
conjectures, and generally open the field up for other investigators
to do more work in it.\footnote{\nocite{earman-primer-determ}This
  paper, in other words, has as its goal a more modest version of that
  of Earman's wonderful book \emph{A Primer on Determinism}, to which
  the name of this paper is an homage.}

\section{The Standard Energy Conditions}
\label{sec:standard-energy-conds}

\resetsec

There are several different ways to formulate all the energy
conditions standardly deployed in classical general relativity, both
as a group and individually.  I will focus here on three ways of
formulating them as a group, what one may think of as the
\emph{geometric}, the \emph{physical} and the \emph{effective} ways,
and will for a few of them discuss as well alternative individual
formulations according to the geometric and physical ways, as they
variously allow different insights into the character of the
conditions.\footnote{In this section, aside from a few idiosyncracies,
  such as my classification of different types of formulation, I
  follow in part the exposition of
  \citeN[ch.~12]{visser-lorentz-worms} and in part that of
  \citeN[\S2.5 and \S2.8]{malament-fnds-gr-ngt} for the formulations
  of the conditions themselves.  See \citeN{curiel-econds-cosmo} for
  another formulation of them, based on the scale factor $a(t)$ in
  generic cosmological models, and discussion thereof.}  The geometric
and physical ways are easy to characterize: for the former, one writes
down formal conditions expressed by use only of the value of a purely
geometric tensor (such as the Ricci or Weyl tensor), perhaps as it is
required to stand in relation to a fixed family of vectors or other
tensors; for the latter, one writes down formal conditions expressed
by use only of the value of the stress-energy tensor itself, perhaps
as it is required to stand in relation to a fixed family of vectors or
other tensors.  In every case, the physical formulation is logically
equivalent to the geometric formulation if the Einstein field equation
is assumed to hold.

The effective way requires a bit of groundwork to explain.  According
to a useful classification of stress-energy tensors given by
\citeN[p.~89]{hawking-ellis-lrg-scl-struc-st}, a stress-energy tensor
is said to be of type \textsc{i} if at every point there is a $1 + 3$
orthonormal frame with respect to which it is diagonal,
\emph{i}.\emph{e}., if its only non-zero components as computed in the
given frame are on the diagonal in its matrix form.  In this case, it
is natural to interpret the timelike-timelike component as the
ordinary (mass-)energy density $\rho$ as represented in the given
frame, and the three spacelike-spacelike components to be the three
principal pressures $p_{\hat{\mu}}$ ($\hat{\mu} \in \{1, \, 2, \,
3\}$) as represented in the frame, to be understood by analogy with
the case of a fluid or an elastic body.  The effective formulation of
an energy condition can then be stated as a quantitative relation
among $\rho$ and $p_{\hat{\mu}}$.  Since all known ``physically
reasonable'' classical fields (and indeed many unreasonable ones) have
associated stress-energy tensors of type \textsc{i}, this is no
serious restriction.\footnote{The one possible exception to this claim
  is a null fluid, which has a stress-energy tensor of the form
  $T_{ab} = \rho k_a k_b + p_1 x_a x_b + p_2 y_a y_b$, where $k^a$ is
  null and $x^a$ and $y^a$ are unit spacelike vectors orthogonal to
  $k^a$ and to each other.}  Thus, except for one special case to be
discussed below, the effective formulation should be understood to be
in all ways physically equivalent to the geometric and the physical
formulations, under the assumption that the Einstein field equation
holds, and matter is not too exotic.  Under that assumption, the
effective formulations become especially useful in cosmological
investigations, since the matter fields in standard cosmological
models, the FLRW spacetimes, can always be thought of as
fluids.

It will be convenient to break the conditions up into two further
classes, those (\emph{pointilliste}) that constrain behavior at
individual points and those (\emph{impressionist}) that constrain
average behavior over spacetime regions.  I shall first list the
definitions of all the former, then discuss the significance and
interpretation of each as it will be useful to have them all in hand
at once for the purposes of comparison, then do the same for the
latter class.

\subsection{Pointilliste Energy Conditions}
\label{sec:point-energy-cond}

\resetsub

\begin{description}
  \setlength{\itemsep}{0in}
    \item[null energy condition (NEC)] \hspace*{1em}
  \begin{description} 
      \item[geometric] for any null vector $k^a$, $R_{mn} k^m k^n \geq
    0$
      \item[physical] for any null vector $k^a$, $T_{mn} k^m k^n \geq
    0$
      \item[effective] for each $\hat{\mu}$, $\rho + p_{\hat{\mu}} \geq
    0$
  \end{description}
    \item[weak energy condition (WEC)] \hspace*{1em}
  \begin{description}
      \item[geometric] for any timelike vector $\xi^a$, $G_{mn} \xi^m
    \xi^n \geq 0$
      \item[physical] for any timelike vector $\xi^a$, $T_{mn} \xi^m
    \xi^n \geq 0$
      \item[effective] $\rho \geq 0$, and for each $\hat{\mu}$, $\rho
    + p_{\hat{\mu}} \geq 0$
  \end{description}
    \item[strong energy condition (SEC)] \hspace*{1em}
  \begin{description}
      \item[geometric] for any timelike vector $\xi^a$, $R_{mn} \xi^m
    \xi^n \geq 0$
      \item[physical] for any timelike vector $\xi^a$, $(T_{mn} -
    \half T g_{mn}) \xi^m \xi^n \geq 0$
      \item[effective] $\rho + \sum_{\hat{\mu}} p_{\hat{\mu}} \geq 0$, and
    for each $\hat{\mu}$, $\rho + p_{\hat{\mu}} \geq 0$
  \end{description}
    \item[dominant energy condition (DEC)] \hspace*{1em}
  \begin{description}
      \item[geometric]  \hspace*{1em}
    \begin{enumerate}
        \item for any timelike vector $\xi^a$, $G_{mn} \xi^m \xi^n
      \geq 0$, and $G^a {}_n \xi^n$ is causal 
        \item for any two co-oriented timelike vectors $\xi^a$ and
      $\eta^a$, $G_{mn} \xi^m \eta^n \geq 0$
    \end{enumerate}
      \item[physical] \hspace*{1em}
    \begin{enumerate}
        \item for any timelike vector $\xi^a$, $T_{mn} \xi^m \xi^n
      \geq 0$, and $T^a {}_n \xi^n$ is causal
        \item for any two co-oriented timelike vectors $\xi^a$ and
      $\eta^a$, $T_{mn} \xi^m \eta^n \geq 0$
    \end{enumerate}
      \item[effective] $\rho \geq 0$, and for each $\hat{\mu}$,
    $|p_{\hat{\mu}}| \leq \rho$
  \end{description}
    \item[strengthened dominant energy condition (SDEC)] \hspace*{1em}
  \begin{description}
      \item[geometric] \hspace*{1em}
    \begin{enumerate}
        \item for any timelike vector $\xi^a$, $G_{mn} \xi^m \xi^n
      \geq 0$, and, if $R_{ab} \neq 0$, then $G^a {}_n \xi^n$ is
      timelike
        \item either $G_{ab} = 0$, or, given any two co-oriented
      causal vectors $\xi^a$ and $\eta^a$,

      $G_{mn} \xi^m \eta^n > 0$
    \end{enumerate}
      \item[physical]  \hspace*{1em}
    \begin{enumerate}
        \item for any timelike vector $\xi^a$, $T_{mn} \xi^m \xi^n
      \geq 0$, and, if $T_{ab} \neq 0$, then $T^a {}_n \xi^n$ is
      timelike
        \item either $T_{ab} = 0$, or, given any two co-oriented
      causal vectors $\xi^a$ and $\eta^a$, 

      $T_{mn} \xi^m \eta^n > 0$
    \end{enumerate}
      \item[effective] $\rho \geq 0$, and for each $\hat{\mu}$,
    $|p_{\hat{\mu}}| \leq \rho$
  \end{description}
\end{description}
(It is not an error that the given effective forms of the DEC and the
SDEC are identical; this is the one special case, mentioned above, in
which the effective form of the energy condition diverges from the
geometrical and physical forms.  Of course, it is the case that when
one restricts attention to stress-energy tensors of type \textsc{i},
then the geometrical and physical forms of the DEC and SDEC also
coincide.)  I first sketch the most more or less straightforward
interpretations of the conditions, before discussing problems with
those interpretations.

The idea of average radial acceleration (explained in detail in the
technical appendix \S\ref{sec:tech-app-ara} below) offers one
seemingly promising route toward an interpretation of the geometric
and physical forms of the NEC{}.  Roughly speaking, the average radial
acceleration of a geodesic $\gamma$ at a point $p$ is the averaged
magnitude of the acceleration of neighboring geodesics relative to
$\gamma$ in directions orthogonal to $\gamma$.  If the average radial
acceleration is negative, then this represents the fact that, again
roughly speaking, neighboring geodesics tend to fall inwards towards
$\gamma$ at $p$.  Thus, according to equation~\eqref{eq:ara-ricci},
the geometric form of the NEC{} requires that null geodesic
congruences tend to be convergent in sufficiently small neighborhoods
of every spacetime point (or at least not divergent).  Assuming the
Einstein field equation, the physical interpretation of negative
average radial acceleration for causal geodesics is that, again
roughly speaking, the ``gravitational field'' generated by the ambient
stress-energy is ``attractive''.  Thus, according to
equation~\eqref{eq:ara-stress-en}, the interpretation of the physical
form is that particles following null geodesics will observe that
``gravity'' tends locally to be ``attractive'' (or at least not
repulsive) when acting on nearby particles also following null
geodesics.  Another possible interpretation of the physical form of
the NEC{} is that an observer traversing a null curve will measure the
ambient (ordinary) energy density to be positive.

The interpretation of the effective form of the NEC{} is that the
natural measure either of mass-energy or of pressure in any given
spacelike direction can be negative as determined by an observer
traversing a null curve, but not both, and, if either is negative, it
must be less so than the other is positive.  In so far as one may
think of pressure as a momentum flux, therefore, and so equivalent
relativistically to a mass-energy flow, the effective form requires
that ordinary mass-energy density at any point cannot be negatively
dominated by momentum fluxes in any given spacelike direction as
determined by an observer traversing a null curve: one cannot
indefinitely ``mine'' energy from a system by subjecting it to
negative momentum flux.

The interpretation of the physical form of the WEC is straightforward:
the (ordinary) total energy density of all matter fields, as measured
in a natural way by any observer traversing a timelike curve, is never
negative.  The interpretation of the geometric form is not
straightforward.  Indeed, I know of no simple, intuitive picture that
captures the geometrical significance of the condition.\footnote{It
  has gone oddly unremarked in the physics and philosophy literatures,
  but is surely worth puzzling over, that the Einstein tensor itself,
  the fundamental constituent of the Einstein field equation, has no
  simple, natural geometrical interpretation, in the way,
  \emph{e}.\emph{g}., that the Riemann tensor can naturally be thought
  of as a measure of geodesic deviation.  Perhaps one could try to use
  the Bianchi identity to construct a geometric interpretation for
  $G_{ab}$, or the Lanczos tensor (see
  footnote~\ref{fn:lanczos-tens}), but it is not immediately obvious
  to me what such a thing would look like, if possible.  One can give
  a geometrical interpretation of $G_{ab}$ at a point by considering
  all unit timelike vectors at the point; the Einstein tensor can then
  be reconstructed by defining it to be the unique symmetric two-index
  covariant tensor at that point such that its double contraction with
  every unit timelike vector equals minus one-half the spatial scalar
  curvature of the spacelike hypersurface with vanishing extrinsic
  curvature orthogonal to the given vector.  (See \citeNP[ch.~2,
  \S7]{malament-fnds-gr-ngt}.)  This may be only a matter of taste,
  but I find this interpretation somewhat obscure, certainly not
  simple and natural, in large part because it relies on structure in
  a family of three-dimensional objects to fix the meaning of a
  four-dimensional object.}  The interpretation of the effective form
is similar to that for the NEC{}.  Ordinary mass-energy density must
be non-negative as experienced by any observer traversing a timelike
curve, and the pressure in any given spacelike direction can never be
so negative as to dominate that value.\footnote{Classically, some
  fluids such as water are known to exhibit negative pressures in some
  regimes as measured by observers traversing timelike curves
  (\emph{viz}., us), but these negative pressures are never large
  enough to dominate the fluid's mass-energy.}

It is easy to see, by considerations of continuity, that the WEC
implies the NEC.  \citeN{tipler-econds-st-sings} proved two
propositions that give some insight into the relation between the
NEC{} and the WEC, and into the character of the WEC itself.  He first
showed that, in a natural sense, the WEC is the weakest local energy
condition one can define.  (``Local'' here means something like:
holding at a point, for all observers.)  In particular, he proved the
following: if $T_{mn} \xi^m \xi^n$ is finitely bounded from below for
all timelike $\xi^a$, \emph{i}.\emph{e}., if there exists a $b > 0$
such that $T_{mn} \xi^m \xi^n \geq -b$ for all timelike $\xi^a$, then
WEC holds (\emph{i}.\emph{e}., the supremum of all such $b$ is 0).  He
next proved that one cannot do better by imposing further natural
constraints on the condition: if $T_{mn} \xi^m \xi^n$ is finitely
bounded from below for all unit timelike $\xi^a$, and $T_{ab}$ is of
type \textsc{i}, then the NEC{} holds.  The effective form of the WEC,
therefore, is in fact essentially equivalent to the NEC.  Thus, the
WEC is not the weakest condition in a logical sense one can impose,
but it is the weakest in a loose, physical sense: one cannot do better
by imposing further natural restrictions.

The interpretation of the geometric form of the SEC is similar to that
of the NEC.  According to equation~\eqref{eq:ara-ricci}, the geometric
form of the SEC requires that timelike geodesic congruences tend to be
convergent in sufficiently small neighborhoods of every spacetime
point.  This implies that congruences of null geodesics at that point
are also convergent.  Similarly, according to
equation~\eqref{eq:ara-stress-en}, the interpretation of the physical
form is that observers following timelike geodesics will see that
``gravity'' tends locally to be ``attractive'' in its action on stuff
following both timelike and null geodesics.\footnote{This explication
  of the physical form of the SEC clearly illustrates why it is
  problematic to try to think of general relativity as a theory of
  ``gravity'', in the sense of a force exerted on a body: for bodies
  traversing non-geodetic curves, that is, for bodies experiencing
  non-trivial acceleration, one has no natural way to judge whether
  ``the force of gravity'' is acting attractively or repulsively, not
  even when one fixes a standard of rest (a fiducial body traversing a
  timelike geodesic).  \emph{Pace} particle physicists, general
  relativity simply cannot be comprehended as a theory describing a
  dynamical ``force'' at all.}  The effective form of the SEC has part
of its interpretation the same as that of the WEC, \emph{viz}.,
ordinary mass-energy density at any point cannot be negatively
dominated by momentum fluxes in any given spacelike direction as
determined by an observer traversing a timelike curve.  It also says,
however, that ordinary mass-energy density cannot be negatively
dominated by the sum of the individual pressures (momentum fluxes) at
any point, as determined by an observer traversing a timelike curve.
I know of no compelling eludication of the physical content of that
relation.  The SEC does not imply the WEC, for the SEC can be
satisfied even if the ordinary mass-density is negative.  The SEC
does, however, imply the NEC.  

As for the WEC, the interpretations of the geometrical forms of the
DEC and the SDEC are not clear.  The interpretations of their physical
forms are apparent: every timelike observer will measure ordinary
mass-energy density to be non-negative, and will also measure total
flux of energy-momentum to be causal, with the flow oriented in the
same direction as the observer's proper time.  The SDEC, as the name
suggests, is slightly stronger in that it requires energy-momentum
flux as measured by any timelike observer to be strictly timelike for
non-trivial stress-energy distributions.  The DEC (and \emph{a
  fortiori} the SDEC) are, therefore, standardly taken to rule out
``superluminal propagation of stress-energy''.  (See,
\emph{e}.\emph{g}., the exemplary remarks of
\citeNP[p.~219]{wald-gr}.)  As already noted, the effective forms of
the DEC and SDEC are identical.  Their interpretation, besides the
now-familiar demand that locally measured energy density be
non-negative, is that pressures be strictly bounded both above and
below by the energy density.  This means that the effective fluid can
be neither too ``stiff'' nor too ``lax'', but must lie in a middling
Goldilocks regime.\footnote{\label{fn:w-index}See
  \citeN{curiel-econds-cosmo} for a discussion of the consequences of
  allowing the effective fluid to be too lax, which is to say,
  allowing the barotropic index $w$ to be less than -1, in the context
  of cosmology.  $\displaystyle w \coloneq \frac{p}{\rho}$, and so is
  a useful measure of the ``stiffness'' of whatever (nearly)
  homogeneous, isotropic stuff is used in cosmological models to fill
  spacetime.}  The second given geometric and physical forms of the
SDEC make it manifest that the SDEC is in fact logically stronger than
the DEC.  Of course, any $T_{ab}$ that satisfied the DEC but violated
the SDEC would have to be not of Hawking-Ellis type~\textsc{i}, for it
is only in that case that the two come apart.  Clearly, the SDEC
implies the DEC, which implies the WEC.  

Before turning to discuss the so-called impressionist energy
conditions, I briefly discuss a few problems with the interpretations
I have sketched of the pointilliste conditions.  The interpretations
of the geometrical and physical forms of the NEC{} based on average
radial acceleration is undermined by the fact that convergence of null
geodesics at a point does not in general imply convergence of all
timelike geodesics at that point.  This is why I hedged the proposed
interpretations with slippery terms like `tends to': even if the NEC{}
is satisfied at a point, an observer traversing a timelike geodesic
may still see ``gravity acting repulsively'' in a small neighborhood.
The existence of a negative cosmological constant is a case in which
NEC{} is satisfied, but, by the failure of the SEC, there is still
divergence of timelike geodesics: ``gravity acts repulsively'' on
matter following timelike geodesics, even though it ``acts
attractively'' on stuff following null geodesics.  The other proposed
interpretation of the physical form of the NEC---that observers
traversing null curves will measure non-negative energy
density---suffers from the fact that it is difficult to see what
physical sense can be made of the idea of an observer traveling at the
speed of light making (ordinary) energy measurements.  One cannot try
to ameliorate this problem by positing that the condition means only
that a physical system traversing a null curve will ``experience''
only non-negative energy densities in its couplings with other
systems, irrespective of whether it is an observer making
measurements: ordinary energy density is not an observer-independent
quantity, and so it can mediate no physical interaction in any way
with intrinsic physical significance.  No physical system will
``experience'' ordinary energy density at
all.\footnote{\label{fn:enrgy-dens-noninvar}We decompose $T_{ab}$ into
  energy density, momentum flux and stress in our
  \emph{representations} of our experiments, for various pragmatic and
  psychological reasons; the decomposition represents nothing of
  intrinsic physical significance about the world.  This fact perhaps
  lies at the root of most if not all the difficulties and puzzles
  that plague the energy conditions, especially why they do not seem
  to be derivable from other fundamental principles.  Of course, this
  fact also makes it even more puzzling that they should have such
  profound, physically significant consequences as they do.  What is
  going on here?.}

The interpretation of the effective form of the NEC{} suffers the same
difficulty: what physical content does it have to compare the
magnitude of ordinary energy density and that of momentum flux in a
given spacelike direction, as determined by an observer traversing a
\emph{null} curve?  There is an even more serious problem here,
though, which the effective form makes particularly clear, showing the
limitations of the physical significance of the NEC.  Assuming a well
behaved barotropic equation of state for the effective fluid described
by the stress-energy tensor, \emph{i}.\emph{e}., a fixed relation
$\rho (p)$ expressing $\rho$ as an invertible function of the single
isotropic pressure $p$, the speed of sound is defined by
$\displaystyle c_s^2 \coloneq \frac{\text{d} p}{\text{d} \rho}$.  It
should be clear that the NEC{} does \emph{not} require that $c_s \leq
1$; in other words, stuff can satisfy the NEC{} while still permitting
superluminal propagation of physically significant structure.  It is
thus unclear in the end what real physical significance the
requirement that mass-energy density not be negatively dominated by
momentum fluxes has.

The problems with the effective interpretation of the WEC are much the
same as for the NEC: it is not clear what physical significance the
given relations among energy density and pressure can have when they
permit superluminal propagation of physical structure.  The fact that
the WEC requires energy density always to be positive may make one at
first glance think that it will be violated in the ergosphere of a
Kerr black hole, where, as is well known, ordinary systems can have in
a natural sense negative energy
\cite{penrose-grav-coll-role-gr,penrose-floyd-extract-rot-enrgy-bh}.
In fact, though, there is an equivocation on `energy' here that points
to a subtle and important point.  The energy that can be negative near
a Kerr black hole is the energy defined by the stationary Killing
field of the spacetime, not the ordinary energy density as measured by
any observer using tools at rest with respect to herself.  (Because
the stationary Killing field is spacelike in the ergosphere, no
observer can have its orbits as worldine.)  Now, as I remarked in
footnote~\ref{fn:enrgy-dens-noninvar}, ordinary energy density, not
being an observer-independent quantity, is not a particularly natural
concept in general relativity.  The energy defined by a stationary
Killing field, however, is observer-independent and so has \emph{prima
  facie} physical significance, even more so given that it obeys both
a local and an integral conservation law.  Why is it not troubling
that \emph{this} quantity, a manifestly deep and important one, can be
negative, whereas the negativity of the observer-dependent ordinary
energy density throws us into fits?  Why do we depend so strongly on
conditions formulated using quantities that, under their standard
physical interpretation, are not observer-independent, especially when
proving results about observer-independent quantities and structures,
such as event horizons, that are?  I don't know.  Perhaps the lesson
here is that the geometric form of the energy conditions are the ones
to be thought of as fundamental, in so far as they rely for their
statement and interpretation only on invariant, geometrical structures
and concepts.  It would then be an interesting problem why in the
context of some theories, such as general relativity, the physical
interpretation of the conditions turns out to have questionable
significance.  Perhaps this is telling us to look for theories in
which these important geometric conditions have physically significant
interpretations.  I will return to discuss this question in
\S\ref{sec:constraints-char-st-theors}.

With regard to the SEC, because the convergence of all timelike
geodesics at a point does imply the convergence of null geodesics
there, the proposed interpretations of its geometric and physical
forms, that ``gravity tends to be attractive'', are on firmer ground
than for the NEC.  There is still a problem, though, even here.
Averaged radial acceleration is, after all, only an average,
factitious quantity.  That it be negative does not say that individual
freely falling ordinary bodies cannot in fact accelerate away from
each other for no apparent reason, only that, on average, they do not
do so.  Thus, the idea that average geodetic convergence should be
thought of as a representation of the attractiveness of gravity is
dicey at best.  And, again, there is the issue that this condition
says nothing at all about the ``effect of gravity'' on bodies
accelerating under the action of other forces.

The DEC (and \emph{a fortiori} the SDEC) are standardly taken to rule
out ``superluminal propagation of stress-energy''.  Once again,
however, it is clear that the DEC does not preclude superluminal
speeds of sound for fields, so it is not clear what work the
prohibition on superluminal propagation of stress-energy is doing.
Even if we put that point aside, though, there are other problems, as
\citeN{earman-no-superlum-propag} argues, claiming the DEC ought not
be interpreted as prohibiting superluminal propagation of
stress-energy.  His argument goes in two steps.  He first argues for
the positive conclusion that the proper way to conceive of a
prohibition on superluminal propagation is the existence of a well
posed (in the sense of Hadamard) initial-value formulation for all
fields on spacetime.  Then, based on \citeN{geroch-faster-light}, he
shows that physical systems can have well posed initial-value
formulations even when the DEC is violated.  Earman's arguments are
buttressed by a recent argument due to
\citeN{wong-reg-hyperbol-dom-econd-caus-lagn}.  As Wong notes (along
with Earman), the evidence almost always cited in support of the idea
that DEC prohibits superluminal propagation of stress-energy is the
theorem that states that, if a covariantly divergence-free $T_{ab}$ is
required to satisfy the DEC and it vanishes on a closed, achronal set,
then it vanishes in the domain of dependence of that set
\cite{hawking-cons-matt-gr,hawking-ellis-lrg-scl-struc-st}.\footnote{A
  region of spacetime is achronal if no two of its points stand in
  timelike relation to each other.  The domain of dependence $D
  (\Sigma)$ of a closed achronal set $\Sigma$ is the collection of all
  points $p$ in spacetime such that every inextendible causal curve
  passing through $p$ intersects $\Sigma$.}  Wong, I think rightly,
points out that this theorem in fact shows only that DEC prohibits
``the edge of a vacuum'' (or vacuum fluctuations, in a quantum
context) from propagating superluminally, not arbitrary stress-energy
distributions.  Given the nonlinearity of the Einstein field equation,
I find it plausible that there may be problems in trying to naively
generalize this result to arbitrary stress-energy tensors, whether
they obey the DEC or not.

Comparing the strengths and weaknesses of the interpretations of the
different forms of the conditions amongst themselves reveals some
interesting questions.  Consider the NEC: on the face of it, the
geometric form has a relatively unproblematic interpretation, whereas
the interpretations of the physical and effective forms are beset with
more serious problems.  The case is just the opposite for the WEC: the
geometric form has no clear interpretation, whereas the physical form
and at least part of the effective form (the positivity of energy
density) are relatively unproblematic.  The DEC occupies yet more
treacherous ground, in so far as the geometric form has no clear
interpretation, the physical interpretation (as Earman's and Wong's
arguments show) is muddled at best, and the effective is only
partially unproblematic.  And yet these statements are, modulo the
assumption of the Einstein field equation, logically equivalent.
Ought unclarity of interpretation of one form push us to question the
seeming clarity of interpretation of other forms?  How can this
happen, that the interpretation of one proposition can be problematic
while the interpretation of a proposition logically equivalent is not
(or, at least, is less so)?  Can we lay all the blame on the
assumption of the Einstein field equation?  I don't think so, for, if
we could, then surely the forms that had interpretive problems would
all be of the same type, but that is not the case here.  Sometimes it
is the geometric that is less problematic, and other times it is the
more problematic.

This is not the place to try to address these questions.  I will
remark only that this topic would provide very rich fodder for an
investigation into the relations between pure geometry and the
physical systems that geometry purports to represent in a given
theory, what must be in place in order to extract physically
significant information from the geometry of those systems, and what
the difference is between having an interpretation of a piece of pure
mathematics and having a physical interpretation of it in the context
of a theory.  I have the sense that it is often a tacit assumption in
philosophical discussions of the meaning of theoretical terms that, if
a mathematical structure has a clear physical interpretation in a
theory, then it itself must have a clear mathematical interpretation
already.  These examples show that this need not be so.  They also
provide interesting case studies of how theoretically equivalent
statements can seemingly have very different physical meanings.

I conclude this section with an observation of what is \emph{not}
here: there are no standard energy conditions based on the Weyl
conformal tensor $C^a {}_{bcd}$ or on the Bel-Robinson tensor
$T_{abcd}$.\footnote{For characterization and discussion of the
  Bel-Robinson tensor and its properties, see
  \citeN{penrose-rindler-spinors-st-1},
  Senovilla~\citeyear{senovilla-superenrgy-tens,senovilla-superergy-tens-apps},
  \citeN{garecki-rmrks-bel-rob-tens} and
  \citeN{gomez-lobo-dynal-laws-supenrgy-gr}.}  I find this odd.  The
standard pointilliste energy conditions do not directly constrain the
behavior of anything one may want to think of as gravitational
stress-energy, and yet one may still want to try to do so.  The
possible need for trying to do so becomes clear when one considers how
strange, even pathological, purely vacuum spacetimes can be, such as
Taub-NUT spacetime and some gravitational plane-wave
spacetimes.\footnote{See, \emph{e}.\emph{g}., \citeN{misner67} and
  \citeN{ellis-schmidt77}, respectively, and \citeN{curiel-sing} for
  further discussion.}  Because the Weyl tensor is not directly
constrained by the stress-energy tensor of matter, in the sense that
it may be non-zero even when $T_{ab}$ is zero, it is often thought to
represent ``purely gravitational'' degress of
freedom.\footnote{\label{fn:lanczos-tens}Still, $C^a {}_{bcd}$ and
  $T_{ab}$ are not entirely independent of each other.  If we define
  the so-called Lanczos tensor
  \begin{equation}
    \begin{split}
      J_{abc} &= \frac{1}{2} \nabla_{[b} R_{a]c} + \frac{1}{6}
      g_{c[a} \nabla_{b]} R \\ 
      &= 4 \pi \nabla_{[b} T_{a]c} - \frac{1}{12} g_{c[b} \nabla
      _{a]} T
    \end{split}
  \end{equation} 
  then the Bianchi identities may be rewritten 
  \[
  \nabla_n C^n {}_{abc} = J_{abc}
  \]
  The similarity of this equation to the sourced Maxwell equation
  suggests regarding the Bianchi identities as field equations for the
  Weyl tensor, specifying how at a point it depends on the
  distribution of matter at nearby points.  (This approach is
  especially useful in the analysis of gravitational radiation; see,
  for example, \citeNP{newman-penrose62}, \citeNP{newman-unti62}, and
  \citeNP{hawking-pertub-expan-uni}.)  Thus, conditions imposed on the
  Weyl tensor might still be plausibly interpretable as energy
  conditions in spacetimes with non-trivial $T_{ab}$.}  The
Bel-Robinson tensor, moreover, may usefully be thought of as a measure
of a kind of ``super-energy'' associated with purely gravitational
phenomena, and directly measures in a precise sense the intensity of
gravitational radiation in infinitesimal regions.  These two tensors,
therefore, would seem perfect candidates to serve as the basis for
conditions that would constrain the behavior of purely gravitational
phenomena and, more particularly, of vacuum spacetimes.  I think it
would be of great interest to investigate whether there are natural
conditions based on these two tensors that would constrain behavior in
vacuum spacetimes so as to rule out such pathologies.  I conjecture
that there are indeed such conditions.  One potentially promising
place to start a search for such conditions might be the Weyl
Curvature Hypothesis of \citeN{penrose-sing-time-asym}, and recent
work attempting to formulate expressions for gravitational entropy
based on these two tensors.\footnote{See, \emph{e}.\emph{g}.,
  \citeN{cotsakis-klaoudatou-cosmo-sings-bel-rob} and
  \citeN{clifton-et-gravl-ent-propos}.}

\subsection{Impressionist Energy Conditions}
\label{sec:impress-energy-cond}

\resetsub

Before exhibiting the impressionist energy conditions, a little
technical background is in order.  If $\gamma$ is a timelike curve,
then it is natural to parametrize the line-integral of a quantity
along $\gamma$ by proper time.  If $\gamma$ is a null curve, however,
one does not have a natural parametrization of it available.  In this
case, it is convenient to use a generalized affine
parameter.\footnote{See, \emph{e}.\emph{g}., \citeN{schmidt71} for a
  definition and discussion.}  The generalized affine parameter is
especially useful in that it does not depend on the basis chosen in
one crucial respect: whether or not the generalized affine parameter
of the curve increases without bound.

In order to express the impressionist conditions in effective form, it
will be convenient to define direction cosines for causal tangent
vectors.  Fix a $1 + 3$ orthonormal frame with respect to which the
stress-energy tensor (assumed, recall, for the effective form, to be
of Hawking-Ellis type \textsc{i}) is diagonal.  Let $k^\mu$ be the
components of the null vector $k^a$ with respect to the fixed frame.
Then define the normalization function $\nu_n$ and the direction
cosines $\cos \alpha_\mu$ so that $\cos \alpha_0 = 1$ and $k^\mu =
\nu_n (k^a) \cos \alpha_\mu$.  Let $\xi^\mu$ be the components of the
timelike vector $\xi^a$ with respect to the fixed frame.  Then define
the normalization function $\nu_t$, the real number $\beta$, and the
direction cosines $\cos \alpha_\mu$ so that $\cos \alpha_0 = 1$,
$\xi^0 = \nu_t (\xi^a) \cos \alpha_0$ and $\xi^{\hat{\mu}} = \nu_t
(\xi^a) \beta \cos \alpha_{\hat{\mu}}$.

Although in principle one could define impressionist energy conditions
based on spacetime regions of any dimension or topology, in practice,
at least in the classical regime, they have all been defined using
curves of various types.  In my exposition of them here, I will give
what is in effect only a template for the ones actually used to prove
theorems, which often qualify the basic template in some way.  I will
explain or at least mention some of those qualifications in my
discussion below in this section, and also in \S\ref{sec:conseq-viol}.
All the impressionist energy conditions based on curves have this in
common: the characteristic property that is postulated is required to
hold on every curve in some fixed class $\Gamma$ of curves on
spacetime.
\begin{description}
  \setlength{\itemsep}{0in}
    \item[averaged null energy condition (ANEC)] \hspace*{1em}
  \begin{description} 
      \item[geometric] for every $\gamma$ in the fixed class of null
    curves $\Gamma$,
    \[
    \int_\gamma R_{mn} k^m k^n \: \text{d} \theta \geq 0
    \]
    where $\gamma$ has tangent vector $k^a$ and $\theta$ is a
    generalized affine parameter along $\gamma$
      \item[physical] for every $\gamma$ in the fixed class of null
    curves $\Gamma$,
    \[
    \int_\gamma T_{mn} k^m k^n \: \text{d} \theta \geq 0
    \]
    where $\gamma$ has tangent vector $k^a$ and $\theta$ is a
    generalized affine parameter along $\gamma$
      \item[effective] for every $\gamma$ in the fixed class of null
    curves $\Gamma$,
    \[
    \int_\gamma \left( \rho + \sum_{\hat{\mu}} p_{\hat{\mu}} \cos^2
      \alpha_{\hat{\mu}} \right) \nu^2_n (k^a) \: \text{d} \theta \geq
    0
    \]
    where $\gamma$ has tangent vector $k^a$ and $\theta$ is a
    generalized affine parameter along $\gamma$
  \end{description}
    \item[averaged weak energy condition (AWEC)] \hspace*{1em}
  \begin{description} 
      \item[geometric] for every $\gamma$ in the fixed class of
    timelike curves $\Gamma$,
    \[
    \int_\gamma G_{mn} \xi^m \xi^n \: \text{d} s \geq 0
    \]
    where $\gamma$ has tangent vector $\xi^a$ and $s$ is proper time
      \item[physical] for every $\gamma$ in the fixed class of
    timelike curves $\Gamma$,
    \[
    \int_\gamma T_{mn} \xi^m \xi^n \: \text{d} s \geq 0
    \]
    where $\gamma$ has tangent vector $\xi^a$ and $s$ is proper time
      \item[effective] for every $\gamma$ in the fixed class of
    timelike curves $\Gamma$,
    \[
    \int_\gamma \left( \rho + \beta^2 \sum_{\hat{\mu}} p_{\hat{\mu}}
      \cos^2 \alpha_{\hat{\mu}} \right) \nu^2_t (\xi^a) \: \text{d}s
    \geq 0
    \]
    where $\gamma$ has tangent vector $\xi^a$ and $s$ is proper time
  \end{description}
    \item[averaged strong energy condition (ASEC)] \hspace*{1em}
  \begin{description} 
      \item[geometric] for every $\gamma$ in the fixed class of
    timelike curves $\Gamma$,
    \[
    \int_\gamma R_{mn} \xi^m \xi^n \: \text{d} s \geq 0
    \]
    where $\gamma$ has tangent vector $\xi^a$ and $s$ is proper time
      \item[physical] for every $\gamma$ in the fixed class of
    timelike curves $\Gamma$,
    \[
    \int_\gamma \left( T_{mn} - \half T g_{mn} \right) \xi^m \xi^n \:
    \text{d} s \geq 0
    \]
    where $\gamma$ has tangent vector $\xi^a$ and $s$ is proper time
      \item[effective] for every $\gamma$ in the fixed class of
    timelike curves $\Gamma$,
    \[
    \int_\gamma \left\{ \left( \rho + \beta^2 \sum_{\hat{\mu}}
        p_{\hat{\mu}} \cos^2 \alpha_{\hat{\mu}} \right) \nu^2_t
      (\xi^a) - \half \xi^n \xi_n \left( \rho - \sum_{\hat{\mu}}
        p_{\hat{\mu}} \right) \right\} \text{d}s \geq 0
    \]
    where $\gamma$ has tangent vector $\xi^a$ and $s$ is proper time
  \end{description}
\end{description}
Before discussing their respective interpretations, a few remarks are
in order.  No reasonable impressionist analogue of either of the
pointilliste dominant conditions are known.\footnote{One could in flat
  spacetimes, and possibly in stationary spacetimes, circumvent the
  obvious problems with formulating a dominant-like impressionist
  energy condition, but, being confined to flat (and possibly
  stationary) spacetimes, such a condition would have little import or
  relevance.}  In practice, one generally requires that $\Gamma$
consist of a suitably large family of inextendible geodesics of the
appropriate type.  For the ANEC, if $\Gamma$ consists of null
geodesics, then one can replace the generalized affine parameter with
the ordinary affine parameter.  In no case can one allow arbitrary
parametrizations for null curves in the defining integral, as that
would simply reduce the ANEC to the NEC.  If one further requires for
the ANEC that the curves in $\Gamma$ be achronal, then the condition
is often called the `averaged achronal null energy condition' (AANEC).
For the AWEC, if $\Gamma$ contains enough timelike geodesics and the
spacetime is well behaved, then there may be null geodesics that are
limit curves of sub-families of $\Gamma$; in this case, the relevant
characteristic integral will be non-negative for those null geodesics,
and the AWEC with the fixed $\Gamma$ can be said to imply the ANEC for
the family of limiting null geodesics.  Even in well behaved
spacetimes, however, there may be null geodesics that are not the
limit of any family of timelike geodesics, so in general the AWEC does
not imply the ANEC.  The ASEC does not imply either the AWEC or the
ANEC.  Clearly, the NEC, WEC and SEC respectively imply the ANEC, AWEC
and ASEC.

I am sorry to say the discussion of the possible interpretations of,
or even just motivations for, the standard impressionist energy
conditions is a simple one to have: there are no compelling
geometrical, physical or effective interpretations of these
conditions, not even hand-waving, rough or approximate ones, and no
compelling physical or philosophical motivations for them.

I should perhaps clarify what I mean in claiming that there are no
compelling interpretations or motivations of these conditions.  One
can certainly describe in simple, clear, physical language the sorts
of spacetimes in which they will be satisfied---geodesics experience
more positive than negative energy, the regions in which the
pointilliste conditions are violated are bounded in various ways,
\emph{etc}.---but it is difficult, at best, to understand these
classes of spacetimes as being related in any but accidental ways.
There is nothing principled or lawlike that makes these spacetimes
similar or the same in any deep sense.  It is not easy to imagine
principled conditions one could impose on theories of matter or
fields---say, a form for the Lagrangian, or manifestation of a
symmetry, \emph{etc}.---that would ensure the sort of behavior
captured by the averaged conditions.  This somewhat vague qualm is
substantiated by the ease with which violations of the averaged
conditions can be found, in both the classical and the quantum cases.

More to the point, there is at least one interesting way of making
this vague qualm more precise, that at the same time shows clearly the
artificiality of the impressionist conditions as compared to the
pointilliste conditions: none of the quantities constrained by the
impressionist conditions enter the equations of motion or the field
equations of any known kinds of physical system, and, correlatively,
no couplings between any known kinds of physical system are mediated
by those quantities; the opposite is true for the pointilliste
conditions, whose constrained quantities promiscuously appear in
equations of motion, field equations and couplings for many if not
most known kinds of physical system.  Finally, the restriction to
geodesics has no compelling physical or philosophical basis that I can
see, but appears to be dictated by pragmatic considerations about the
technical tractability of required calculations.

Still, there is more to say about them, even though none has a clear,
principled interpretation or motivation.  These conditions were all
constructed by reverse-engineering---an investigator looked for the
weakest condition she could impose on the averaged behavior of some
quantity depending on curvature or stress-energy in order to derive
the result of interest to her.  (Indeed, I think it is not going too
far to say that many of them represent a case of outright
gerrymandering by the relativity community.\footnote{The only
  physicists I know of to express similar concerns are Visser and
  Barcel\'o
  \cite{visser-barcelo-econds-cosmo-implic,barcelo-visser-twilight-econds};
  indeed they seem to be of the opinion that it is difficult to think
  of all energy conditions, not just the impressionist ones, as little
  more than pragmatically convenient tools whose formulation is driven
  by the technical needs in proving desired theorems.})  Other
researchers were impressed by the weakness of the condition used to
derive the important result, and so picked it up and used it
themselves.  And so the impressionist conditions have been passed down
through the generations of relativists, hand to hand from teacher to
student, powerful, talismanic runes to be brought out and invoked with
precise ceremony on formal occasions, but whose inner significance is
beyond our ken, though their very familiarity often obscures that
fact.\footnote{\skipline[-1]
  \begin{tabbing}
    \hspace*{3em}\=und Das\= und Den\=\kill
    \>und Das und Den, \\
    \>die man schon nicht mehr sah \\
    \>(so t\"aglich waren sie und so gew\"ohnlich), \\
    \>auf einmal anzuschauen: sanft, vers\"ohnlich \\
    \>und wie an einem Anfang und von nah \\
    \>\>--- Rainer Maria Rilke, ``Der Auszug des{} verlorenen Sohnes'' 
  \end{tabbing}}

This is not to say the impressionist energy conditions have no
foundational or physical interest at all.  It is often important to
find the weakest conditions one can to prove theorems whose
conclusions have great weight or significance, such as the
positive-energy theorems or the singularity theorems, if only, for
example, to get as clear as one can on what those conclusions really
depend on.  If one wants to try to extend or modify one's global
theory while ensuring that certain results remain true, for example,
it behooves one to find the weakest conditions from which one can
derive those results.  For we who are interested in the foundations of
the theory in and of itself, however, these impressionist conditions
have little to offer.  Still, because they have been used to prove
deep results of great interest in themselves, it is important to
understand what sorts of system violate and what sorts satisfy the
conditions (which I will discuss in \S\ref{sec:conseq-viol}).

Before moving on, it will be edifying to examine in a little detail
two of the most important technical qualifications made to the
templates I gave of the averaged conditions.
\citeN{tipler-econds-st-sings}, which if my history is not mistaken
was the first use of an averaged condition to prove results of any
depth, required the additional constraint that the characteristic
integral of the averaged condition at issue can equal zero for any
curve only if its integrand (\emph{e}.\emph{g}., $T_{mn} \xi^m \xi^n$
for the physical AWEC) equals zero along the entire curve.  As
\citeN{borde-geo-focus-econds-sings} points out, this constraint
raises problems for the physical plausibility, or at least possible
scope, of the
conditions.\footnote{\citeN{chicone-ehrlich-line-int-ric-curv-conj-pts}
  also pointed out that there were lacun{\ae}\ in Tipler's proofs,
  unrelated to Borde's problems, but that is by the by for our
  purposes, as they also showed how to fix the problems.}  To see the
problem, let us for the sake of definiteness focus attention for the
moment on the physical AWEC.  Then Tipler's constraint rules out cases
where the integral equals zero because the relevant curve passes
endlessly in and out of regions of positive and negative energy
density.  This may not sound so bad at first, until one realizes it
means that, for a spacetime to satisfy the constrained condition,
every curve in the fixed class must eventually traverse only regions
of non-negative energy density, both to the past and the future:
violations of the WEC are to be allowed only in bounded regions in the
interior of spacetime, so to speak.  There seems even less physical
justification for demanding this than for the bare AWEC in the first
place.

To try to address this problem, Borde proposed modifications to the
averaged conditions.  The technical details of his proposals, while
ingenious, are not worth working through for my purposes, as they are
complicated and shed little light on the issues I am discussing.  The
gist of his proposed modifications is this: rather than requiring that
the salient integral equal zero only when its integrand equal zero
everywhere along the curve, we require only that, if the integral
equal zero, then the integrand must be suitably periodic along the
entire curve, \emph{i}.\emph{e}., roughly speaking, that the integrand
visit a neighborhood of zero frequently and that the lengths of the
intervals it spends visiting those neighborhoods not approach zero as
one heads along the curve in either direction.  This allows
application of the averaged condition to situations in which the total
integral may essentially be zero even though there are large and long
violations of the relevant pointilliste condition, such as may occur
for the SEC during inflationary periods of a spacetime.  In this
sense, Borde's modifications do seem an improvement on Tipler's
original version.  One cannot help the feeling though, given the
intricacy and physical opacity of the mathematical machinery required
to formulate Borde's condition, that the problems of physical
interpretation in the sense I sketched above---not having in hand a
principled justification for the condition founded on general,
fundamental principles, but rather only reverse-engineering the
weakest suitable condition one can manage to prove the results one
wants for the particular class of spacetimes one is interested
in---seem perhaps even more severe than before.

\subsection{Appendix: A Failed Attempt to Derive the NEC{} and SEC}
\label{sec:failed-attempt-derive}

\resetsub

It is sometimes claimed (\emph{e}.\emph{g}.,
\citeNP{liu-reboucas-econds-bnds-fTgrav}) that one can derive the
NEC{} and the SEC from the Raychaudhuri equation.  Even though I think
the argument fails, it is of interest to try to pinpoint exactly why
it fails, as it sheds light on why it appears to be difficult to
derive the energy conditions from other fundamental principles (the
difficulty strongly suggested by the lack of convincing derivations).
I will sketch the argument only for the SEC, as that for the NEC{} is
essentially the same, with only a few inessential technical
differences.

Raychaudhuri's equation expresses the rate of change of the scalar
expansion of a congruence of geodesics, as one sweeps along the
congruence, as a function of the expansion itself, of the congruence's
shear and twist tensors, and of the Ricci tensor.  For a congruence of
timelike geodesics with tangent vector $\xi^a$, it takes the form
\begin{equation}
  \label{eq:raychaudhuri-tl}
  \xi^n \nabla_n \theta = - \athird \theta^2 - \sigma_{mn} \sigma^{mn}
  + \omega_{mn} \omega^{mn} - R_{mn} \xi^m \xi^n
\end{equation}
where $\theta$ is the expansion of the congruence, $\sigma_{ab}$ its
shear and $\omega_{ab}$ its twist.\footnote{See, \emph{e}.\emph{g}.,
  \citeN[ch.~9, \S2]{wald-gr} for a derivation and explanation of the
  Raychaudhuri equation for both timelike and null congruences.  There
  is a generalization of the Raychaudhuri equation that treats
  congruences of accelerated curves, but nothing would be gained for
  our purposes by discussing it.}  If the total sum on the righthand
side is negative, then the expansion of the congruence is decreasing
with proper time, \emph{i}.\emph{e}., the geodesics in the congruence
are everywhere converging on each other.  The first term on the
righthand side is manifestly negative, as is the second, since
$\sigma_{ab}$ is spacelike in both indices, and so $\sigma_{mn}
\sigma^{mn} \geq 0$.  For a hypersurface-orthogonal congruence, it
follows directly from Frobenius's Theorem that $\omega_{ab} = 0$.
Thus, if we assume that ``gravity is everywhere attractive'', and we
interpret this to mean that congruences of timelike geodesics which
have vanishing twist should always converge, then, in order to ensure
that the total righthand side of equation~\eqref{eq:raychaudhuri-tl}
is always negative, we require that $R_{mn} \xi^m \xi^n \geq 0$, which
is just the geometrical form of the SEC.

It should be clear why I fail to find the argument compelling.  In
fact, all one can conclude from the demand that the righthand side of
equation~\eqref{eq:raychaudhuri-tl} be non-positive (when $\omega_{ab}
= 0$) is that
\begin{equation}
  \label{eq:raychaudhuri-weak-sec}
  R_{mn} \xi^m \xi^n \geq - \athird \theta^2 - \sigma_{mn} \sigma^{mn}
\end{equation}
everywhere.  Of course, this is not the SEC, but only a weaker form of
the geometric formulation, one that sets a non-constant lower bound on
``how negative'' mass-energy and momentum-energy flux can get
(invoking the physical form of the condition).\footnote{Because the
  lower bound is variable, the propositions of
  \citeN{tipler-econds-st-sings} I discussed in
  \S\ref{sec:point-energy-cond} do not allow one to infer that this
  weaker condition is in fact equivalent to the WEC.}  When one
considers that one can, in every spacetime, find at every point a
congruence of timelike geodesics that has divergent expansion as one
approaches that point, one realizes that the
inequality~\eqref{eq:raychaudhuri-weak-sec} is vacuous, for the
righthand side of the inequality can be made as negative as one likes.
(Proof: in any spacetime, at any point $p$, consider the family of
timelike geodesics defined by the family of unit, past-directed,
timelike vectors at $p$, parametrized by proper time so that each
geodesic's parameter has the value 0 at $p$; there will be some real
number $\epsilon$ such that the class of geodetic-segments defined by
considering all geodesics in the family for proper-time values in the
open interval $(-\epsilon, \, 0)$ defines a proper congruence; that
congruence will have divergent expansion along all its members as one
approaches proper time 0, \emph{i}.\emph{e}., the point $p$, as can be
seen by the fact that any spacelike volume swept along the flow of the
congruence towards $p$ will converge to 0.)

The heart of the problem should now be clear.  Geodesic congruences
are a dime a dozen.  You can't throw a rock in a relativistic
spacetime without hitting a zillion of them, most of them having no
intrinsic physical significance.  Because the pointilliste energy
conditions, moreover, constrain the behavior of curvature terms only
at individual points, and that by reference to all timelike or null
(or both) vectors at those points, one can always find geodesic
congruences that are as badly behaved as wants, in just about any way
one wants to make that idea precise, with respect to how various
measures of curvature evolve along the congruences.  Nonetheless,
geodesic congruences seem to be about the only structure one has
naturally available to work with, if one wants to try to constrain the
behavior of curvature as measured by the contractions of curvature
tensors with causal vectors.  So long as one wants to work with
geodesic congruences, therefore, it seems one must find some way to
restrict the class one allows as relevant to those that are
``physically significant'' in some important and clear way.  I know of
no way to try to address that problem.  Of course, one could always
try to work with structures other than geodesic congruences, but,
again, I know of no other natural candidates to try to use to
constrain the behavior of measures of curvature, given the typical
form of the energy conditions.

Even if one could find natural, compelling ways to restrict attention
to a privileged class of congruences in such a way as to resolve the
technical problems I raised for this kind of argument, there would
still be interpretative problems with this kind of argument.  As I
discussed at the end of \S\ref{sec:point-energy-cond} above, I do not
find it convincing to interpret the fact that causal congruences are
convergent as a representation of the idea that ``gravity is
attractive''.  Without that interpretation, however, one has little
motivation for invoking Raychaudhuri's equation in the first place
without ancillary physical justification.

\subsection{Appendix: Very Recent Work}
\label{sec:very-recent-work}

\resetsub

Very recently, \citeN{abreu-et-ent-bnds-w-param} introduced a new
classical energy condition:
\begin{description}
  \setlength{\itemsep}{0in}
    \item[flux energy condition (FEC)] \hspace*{1em}
  \begin{description} 
    \setlength{\itemsep}{0in}
      \item[geometric] \hspace*{1em}
    \begin{enumerate}
        \item for any timelike vector $\xi^a$, $G^a {}_n \xi^n$ is
      causal
        \item for any timelike vector $\xi^a$, $G^m {}_r G_{ms} \xi^r
      \xi^s \geq 0$
    \end{enumerate}
      \item[physical]  \hspace*{1em}
    \begin{enumerate}
        \item for any timelike vector $\xi^a$, $T^a {}_n \xi^n$ is
      causal
        \item for any timelike vector $\xi^a$, $T^m {}_r T_{ms} \xi^r
      \xi^s \geq 0$
    \end{enumerate}
      \item[effective] for each $\hat{\mu}$, $\rho^2 \geq
    p^2_{\hat{\mu}}$
  \end{description}
\end{description}
There is, as is to be expected, no simple interpretation of its
geometric form.  The simplest interpretation of its physical form is
that the total flux of energy-momentum as measured by any timelike
observer is always causal, albeit the temporal direction of the flux
is not restricted.  Because isotropic tachyonic gases always satisfy
$\rho < \frac{1}{3} p$, with weaker bounds for anisotropic tachyonic
material, the effective form may be interpreted as ruling out the
possibility of tachyonic matter.  Otherwise, I know of no compelling
interpretation of it, as it allows energy density to be unboundedly
negative, so long as the absolute value of pressure is not too great.

\citeN{abreu-et-ent-bnds-w-param} argue that the FEC gives better
support to the claim that the cosmological equation-of-state parameter
$w$ (the so-called barotropic index---see footnote~\ref{fn:w-index})
must be $\leq 1$, and so better substantiates arguments in favor of
entropy bounds they give based on that assumption.  Mart\'in-Moruno
and
Visser~\citeyear{martin-moruno-visser-class-q-econds-vac,martin-moruno-visser-semiclass-econds-vac}
investigated its properties and proposed a quantum analogue of it,
which, they claim, works in several respects better than the standard
quantum energy conditions.\footnote{See \citeN{curiel-econds-qftcst}
  for extended discussion of energy conditions in quantum field-theory
  on curved spacetime.}  The FEC, therefore, shows \emph{prima facie}
promise as being of real physical interest.  It is, moreover,
manifestly weaker than all the other standard energy conditions, as
its characteristic non-linearity (most easily seen in the second given
articulations of its geometric and physical forms, and in its
effective form) ensures that essentially no limit is placed on the
possible negativity of the ordinary mass-energy of matter.  If,
therefore, it bears out its promise for leading to, or at least
supporting, results of interest, it would be a great improvement on
the standard energy conditions.  Because, however, its properties and
consequences are virtually unknown as compared to the standard
conditions, I shall not discuss it further.

Even more recently, \citeN{martin-moruno-visser-semiclass-econds-vac}
proposed two more energy conditions, the determinant energy condition
(DETEC) and the trace-of-square energy condition (TOSEC), and also
proposed quantum analogues for them.  Again, these energy conditions
seem \emph{prima facie} interesting, but even less work has been done
on and with them than the FEC, so I shall not discuss them here
either.

\subsection{Technical Appendix: Average Radial Acceleration}
\label{sec:tech-app-ara}

\resetsub

To characterize the idea of the average radial acceleration of a
causal geodesic,\footnote{I follow the exposition of
  \citeN[\S2.7]{malament-fnds-gr-ngt}, with a few emendations.}  let
$\xi^a$ be a future-directed causal vector field whose integral curves
$\gamma$ are affinely parametrized geodesics.  If $\gamma$ is
timelike, then assume $\xi^a$ to be unit.  Let $\lambda^a$ be a vector
field on $\gamma$ such that at one point $\lambda^n \xi_n = 0$ and
$\pounds_\xi \lambda^a = 0$.  (Note that if $\xi^a$ is null, then
$\lambda^a$ may be proportional to $\xi^a$; otherwise it must be
spacelike.)  Then automatically $\lambda^n \xi_n = 0$ at all points of
$\gamma$.  $\lambda^a$ is usefully thought of as a ``connecting
field'' that joins the image of $\gamma$ to the image of another,
``infinitesimally close'' integral curve of $\xi^a$.  Then $\xi^m
\nabla_m ( \xi^n \nabla_n \lambda^a)$ represents the acceleration of
that neighboring geodesic relative to $\gamma$.  According to the
equation of geodesic deviation,
\[
\xi^m \nabla_m ( \xi^n \nabla_n \lambda^a) = R^a {}_{mnr} \xi^m
\lambda^n \xi^r
\]

Now, fix an orthonormal triad-field $\{ \stackrel{\mu}{\lambda} \!\!^a
\}_{\mu \in \{1, \, 2, \, 3\}}$) along $\gamma$ such that each
$\stackrel{\mu}{\lambda} \!\!^a$ forms a connecting (relative
acceleration) field along $\gamma$.  The magnitude of the radial
component of the relative acceleration in the \nth{$\mu$} direction
then is $- \stackrel{\mu}{\lambda} \hspace{-.5ex}_r \xi^m \nabla_m (
\xi^n \nabla_n \stackrel{\mu}{\lambda} \!\!^r)$.  Fix a point $p \in
\gamma$.  The \emph{average radial acceleration} $A_r$ of $\gamma$ at
$p$ is defined to be
\[
A_r \coloneq - \frac{1}{k} \sum_\mu \stackrel{\mu}{\lambda}
\hspace{-.5ex}_r \xi^m \nabla_m ( \xi^n \nabla_n
\stackrel{\mu}{\lambda} \!\!^r)
\]
where $k$ is 3 if $\xi^a$ is timelike and 2 if null.  It is
straightforward to verify that the average radial acceleration is
independent of the choice of orthonormal triad, so it encodes a
quantity of intrinsic geometric (and physical) significance accruing
to $\xi^a$.  A simple calculation using the equation of geodesic
deviation then shows that 
\begin{equation}
  \label{eq:ara-ricci}
  A_r = - \frac{1}{k} R_{mn} \xi^m \xi^n
\end{equation}
If the Einstein field equation is assumed to hold, it follows that
\begin{equation}
  \label{eq:ara-stress-en}
  A_r = - \frac{8\pi}{k} (T_{mn} - \half T g_{mn}) \xi^m \xi^n
\end{equation}
which reduces in the case of null vectors to
\begin{equation}
  \label{eq:ara-stress-en-null}
  A_r = - 4\pi T_{mn} \xi^m \xi^n
\end{equation}

\section{Consequences and Violations}
\label{sec:conseq-viol}

\resetsec

To study the role of energy conditions in spacetime theories, I will
look at results that do not depend on the imposition of any field
equations (\emph{e}.\emph{g}., the Einstein field-equation) and yet
directly constrain spacetime geometry.  One will often hear the claim
that such-and-such result (\emph{e}.\emph{g}., various singularity
theorems, various versions of the geodesic postulate, the Zeroth Law
of black-hole mechanics, \emph{etc}.\@) that assumes an energy
condition does require the Einstein field equation for its proof, but
one must be careful of such claims.  It is almost always the case, in
fact, that the Einstein field equation is logically independent of the
result (in the strong sense that one can assume the negation of the
Einstein field equation and still derive the result); the Einstein
field equation is used in such cases \emph{only} to provide a physical
interpretation of the assumed energy condition; mathematically, one in
general needs only the geometric form of the condition, which is why I
distinguish the geometric from the physical form.  In this section,
every consequence of the energy conditions I discuss is of this type:
it is logically independent of the Einstein field equation, and relies
on the Einstein field equation only for the physical interpretation of
the assumed geometric energy condition.\footnote{In cosmology, several
  of the most interesting results do require assumption of the
  Einstein field equation.  For this reason, and also because it is
  such a large and rich field on its own, I explore the role and
  character of energy conditions in the context of cosmology at some
  length in \citeN{curiel-econds-cosmo}.}  Many of the violations of
the energy conditions I list here, however, do rely on assuming the
Einstein field equation for their derivation, in so far as they use
the Lagrangian formulation of the relevant forms of matter to derive
the violation, or in so far as they rely on the effective form of the
energy conditions in conjunction with, \emph{e}.\emph{g}., the
Friedmann equations to derive the violation.

I will begin with a list of the consequences of the energy conditions,
\emph{i}.\emph{e}., the results each energy condition is used to
derive, and then discuss the roles the conditions play in the
derivations of those results.  I then list the classical cases in
which energy condition is known to fail, then discuss how the known
failures may or may not undermine our confidence in the
consequences.\footnote{See \citeN{curiel-econds-qftcst} for
  examination of the cases of failure in the quantum regime.}  In
several of the references I give in the list of consequences, no
explicit mention is made of energy conditions, but, if one works
through their arguments, one will see that the relevant energy
condition is indeed being implicitly assumed.  In other works I cite,
an energy condition is explicitly assumed, but in fact, according to
the arguments of those works, either a weaker one is sufficient or a
stronger one is required; in such cases, I cite the result under the
sufficient or required condition.  For almost none of the statements
in the list of consequences is it the case that the energy condition
alone is necessary or sufficient; it is rather that the energy
condition is one assumption among others in the only known way (or
ways) to prove the result.  When I list the same proposition as a
consequence of more than one energy condition (\emph{e}.\emph{g}.,
``prohibition on spatial topology change'' under both WEC and ANEC),
it means that there are different proofs of the statement using
different ancillary assumptions.  In some of the cases of violations I
list, the circumstance or condition possibly leads to a violation of
the germane energy condition; in other cases, it necessarily does so.
I will indicate which is which.  When I list the same type of system
as violating different energy conditions (\emph{e}.\emph{g}., ``big
bang'' singularities for both NEC{} and SEC), it means that different
instances of that type of system violate the different conditions.
When I qualify a spacetime as ``spatially open'' or ``spatially
closed'', it should be understood that the spacetime is globally
hyperbolic and the openness or closedness refers to the topology of
spacelike Cauchy surfaces in a natural slicing of the spacetime.

\subsection{Consequences}
\label{sec:consequences}

\begin{description}
    \item[NEC] \hspace{1em}
  \begin{enumerate}
      \item formation of singularities after gravitational collapse in
    spatially open spacetimes \cite{penrose-grav-coll-st-sings}
      \item formation of singularities in asymptotically flat
    spacetimes with non-simply connected Cauchy surface
    \cite{gannon-sings-nonsimp-conn-sts}
      \item formation of an event horizon after gravitational collapse
    \cite{penrose-grav-coll-st-sings,penrose-struc-st,penrose-grav-coll-role-gr}
      \item trapped and marginally trapped surfaces and apparent
    horizons must be inside asymptotically flat black holes
    \cite{wald-gr}
      \item Hawking's Area Theorem for asymptotically flat black holes
    (Second Law of black-hole mechanics)
    \cite{hawking-gravrad-collidebhs}
      \item the area of a generalized black hole always
    increases\footnote{\label{fn:genl-bhs}\citeN{hayward-genl-laws-bhdyns}
      defines a generalized notion of black hole, one applicable to
      spacetimes that are not asymptotically flat, by the use of what
      he calls ``trapping horizons''.  In the same paper, he shows
      that generalized black holes obey laws analogous to the standard
      Laws of black-hole mechanics.}  (Second Law of generalized
    black-hole mechanics) \cite{hayward-genl-laws-bhdyns}
      \item asymptotically predictable black holes cannot
    bifurcate\footnote{A spacetime is asymptotically predictable if it
      is asymptotically flat, and there is a partial Cauchy surface
      whose boundary is the event horizon, such that future null
      infinity is contained in its future domain of dependence.}
    \cite{wald-gr}
      \item the domain of outer communication of a stationary,
    asymptotically flat spacetime is simply connected, if the domain
    is globally hyperbolic\footnote{The domain of outer communication
      of an asymptotically flat spacetime is, roughly speaking, the
      exterior of the black hole region.  See
      \citeN[\S2.4]{chrusciel-et-stat-bhs-uniq-beyond} for a precise
      definition.  This theorem is similar to the Topological
      Censorship Theorem of \citeN{friedman-et-topol-censor}; see
      footnote~\ref{fn:topol-censor}.}
    \cite{chrusciel-wald-topol-stat-bhs,galloway-topo-dom-outer-comm}
      \item a stationary, asymptotically flat black hole has topology
    $\mathbb{S}^2$, if the domain of outer communication is globally
    hyperbolic and the closure of the black hole is compact
    \cite{galloway-topo-bhs,chrusciel-wald-topol-stat-bhs}
      \item almost all the constituents of the black hole ``No Hair''
    theorem for asymptotically flat black
    holes\footnote{\label{fn:no-hair-thm}The ``No Hair'' theorem
      states that an asymptotically flat, stationary black hole is
      completely characterized by three parameters, \emph{viz}., its
      mass, angular momentum and electric charge.  The proof of this
      theorem logically comprises many steps, each of interest in its
      own right, and historically stretched from the original papers
      of
      Israel~\citeyear{israel-evhoriz-stat-vac-st,israel-evhoriz-stat-elecvac-st}
      to the final results of \citeN{mazur-unique-kerr-newman}.  There
      are too many constituents of the proof to list each
      individually.  A few remaining constituents require the DEC; see
      that list for details.  \citeN{heusler-bh-uniq} provides an
      excellent, relatively up-to-date overview of all the known
      results.  There are analogous ``No Hair'' theorems for the
      generalized black holes of \citeN{hayward-genl-laws-bhdyns}, but
      I will not discuss them.}
    \cite{israel-evhoriz-stat-vac-st,israel-evhoriz-stat-elecvac-st,carter-axisym-bh-2degfree,bekenstein-nonexist-baryon-number-stat-bh,teitelboim-nonmeas-lept-num-bhs,teitelboim-nonmeas-qnums-bhs,wald-emfs-mass-bods,wald-perturb-kerr-bh,mueller-et-bhs-stat-vac-sts,robinson-uniq-kerr,robinson-genl-israels-thm,mazur-unique-kerr-newman}
      \item generalized black holes are regions of ``no escape''
    \cite{hayward-bh-confine}
      \item limits on energy extraction by gravitational radiation
    from colliding asymptotically flat black holes
    \cite{hawking-gravrad-collidebhs}
      \item positivity of ADM mass\footnote{Earlier proofs relied on
      the DEC; see that list for details.}
    \cite{penrose-light-rays-sl-inf-pos-mass-thm,ashtekar-penrose-mass-pos-focus-struc-sl-inf}
      \item The Generalized Second Law of Thermodynamics\footnote{This
      states that the total entropy of the world, \emph{i}.\emph{e}.,
      the entropy of ordinary matter plus the entropy of a black hole
      as measured by its surface area, never decreases. }
    \cite{flanagan-et-class-bousso-bnd-gsl}
      \item Bousso's covariant universal entropy
    bound\footnote{Bousso~\citeyear{bousso-cov-ent-conj,bousso-holo-gen-st},
      clarifying and improving on earlier work by
      Bekenstein~\citeyear{bekenstein-bh-ent,bekenstein-ent-enrgy-bound-bnddsyss,bekenstein-understand-bhent,bekenstein-ent-bnds-bh-remnant},
      \citeN{thooft-quant-space-time},
      \citeN{smolin-bek-bnd-topo-qft-plural-qft},
      \citeN{susskind-world-holo},
      \citeN{corley-jacobson-focus-holog-hypoth}, and
      \citeN{fischler-susskind-holog-cosmo}, conjectured that in any
      spacetime satisfying the DEC the total entropy flux $S_L$
      through any null hypersurface $L$ satisfying some natural
      geometrical conditions must be such that $S_L \leq A/4$, where
      $A$ is a spatial area canonically associated with $L$.
      \citeN{flanagan-et-class-bousso-bnd-gsl} managed to prove the
      bound using the weaker NEC.}
    \cite{flanagan-et-class-bousso-bnd-gsl}
      \item the Shapiro ``time-delay'' is always a delay, never an
    advance\footnote{\label{fn:hyperfast}One can understand this
      result physically as a prohibition on a certain form of
      ``hyper-fast'' travel or communication.  Roughly speaking, this
      is travel in spacetime in which the traveler is measured by
      external observers, in a natural way, to travel faster than the
      speed of light, even though the traveler's worldline is
      everywhere timelike.  It is closely related, though not
      identical, to the idea of traversable wormholes.}
    \cite{visser-et-superlum-cens}
      \item standard formulations of the classical Chronology
    Protection Conjecture\footnote{This states, roughly, that the
      formation of closed timelike curves always requires either the
      presence of singularities or else pathological behavior ``at
      infinity''.}  \cite{hawking-chron-protect-conj}
  \end{enumerate}
    \item[WEC] \hspace{1em}
  \begin{enumerate}
      \item asymptotically flat spacetimes without naked singularities
    are asymptotically predictable \cite{hawking-bhs-in-gr}
      \item asymptotically flat black holes cannot bifurcate
    \cite{hawking-bhs-in-gr}
      \item Third Law of black-hole mechanics\footnote{No physical
      process can reduce the surface gravity of an asymptotically flat
      black hole to zero in a finite amount of time.}
    \cite{israel-3rdlaw-bh}
      \item limits on energy extraction by gravitational radiation
    from asymptotically flat colliding black holes \cite{hawking-bhs-in-gr}
      \item formation of singularities after gravitational collapse in
    spatially open spacetimes
    \cite{geroch-sings,tipler-econds-st-sings}
      \item cosmological singularities in spatially open or flat
    spacetimes \cite{hawking-sing-open-uni,geroch-sings}
      \item cosmological singularities in globally hyperbolic
    spacetimes that are noncompactly regular near
    infinity\footnote{Roughly speaking, a globally hyperbolic
      spacetime is noncompactly regular near infinity if it has a
      (partial) Cauchy surface that is the union of well behaved
      nested sets, each having compact boundary, that are noncompact
      near infinity.}  \cite{gannon-topo-slhs-sing-bhs}
      \item prohibition on spatial topology change
    \cite{geroch-topol-gr,tipler-sings-caus-viol}
      \item geodesic theorems for ``point-particles''
    \cite{eddington-math-theor-rel,einstein-et-grav-eqns-prob-mot}
      \item mass limits for stability of hydrostatic spheres against
    gravitational collapse \cite{bondi-massv-spheres-gr}
      \item some standard forms of the Cosmic Censorship Hypothesis
    \cite{joshi-cosmic-censor}
  \end{enumerate}
    \item[SEC] \hspace{1em}
  \begin{enumerate}
      \item cosmological singularities in spatially closed spacetimes
    \cite{geroch-sings-closed-uni,hawking-sing-cosmo-ii,hawking-sing-cosmo-iii,hawking-ellis-cosmic-bbrad-exist-sings,hawking-penrose-sings-gravcollapse-cosmo,geroch-sings}
      \item cosmological singularities in spatially open spacetimes
    \cite{hawking-sing-cosmo,hawking-sings-in-uni,hawking-ellis-cosmic-bbrad-exist-sings,hawking-penrose-sings-gravcollapse-cosmo,geroch-sings}
      \item cosmological singularities in spacetimes with partial
    Cauchy surfaces
    \cite{hawking-sing-cosmo,hawking-sings-in-uni,hawking-sing-cosmo-iii,hawking-penrose-sings-gravcollapse-cosmo,geroch-sings}
      \item formation of singularities after gravitational collapse in
    spatially closed spacetimes
    \cite{hawking-sing-cosmo-iii,hawking-penrose-sings-gravcollapse-cosmo,geroch-sings}
      \item formation of singularities after gravitational collapse in
    spatially open spacetimes
    \cite{geroch-sings,hawking-penrose-sings-gravcollapse-cosmo}
      \item Lorentzian splitting theorem\footnote{I will give two
      versions of the theorem; see
      \citeN{galloway-horta-reg-lorentz-busemann-fns} for proofs of
      both.  In order to state the first version of the theorem,
      define a \emph{timelike line} to be an inextendible timelike
      geodesic that realizes the supremal Lorentzian distance between
      every two of its points \cite{ehrlich-galloway-tl-lines}.  Then
      the theorem, as first conjectured by \citeN{yau-problems}, is as
      follows: let $(\mathcal{M}, \, g_{ab})$ be a timelike
      geodesically complete spacetime satisfying the SEC; if it
      contains a timelike line, then it is isometric to $(\mathbb{R}
      \times \Sigma, \, t_a t_b - h_{ab})$, where $(\Sigma, \,
      h_{ab})$ is a complete Riemannian manifold and $t^a$ is a
      timelike vector-field in $\mathcal{M}$.  (In particular,
      $(\mathcal{M}, \, g_{ab})$ must be globally hyperbolic and
      static.)

      In order to state the second, we need two more definitions.
      First, the \emph{edge} of an achronal, closed set $\Sigma$ is
      the set of points $p \in \Sigma$ such that every open
      neighborhood of $p$ contains a point $q \ni I^-(p)$, a point $r
      \ni I^+(p)$ and a timelike curve from $q$ to $r$ that does not
      intersect $\Sigma$.  Second, let $\Sigma$ be a non-empty subset
      of spacetime; then a future inextendible causal curve is a
      \emph{future $\Sigma$-ray} if it realizes the supremal
      Lorentzian distance between $\Sigma$ and any of its points lying
      to the future of $\Sigma$
      \cite{galloway-horta-reg-lorentz-busemann-fns}; \emph{mutatis
        mutandis} for a \emph{past $\Sigma$-ray}.  (If $\gamma$ is a
      $\Sigma$-ray, it necessarily intersects $\Sigma$.)  The second
      version of the theorem is as follows: let $(\mathcal{M}, \,
      g_{ab})$ be a spacetime that contains a compact, acausal
      spacelike hypersurface $\Sigma$ without edge and obeys the SEC;
      if it is timelike geodesically complete and contains a future
      $\Sigma$-ray $\gamma$ and a past $\Sigma$-ray $\eta$ such that
      $I^- (\gamma) \cap I^+ (\eta) \neq \emptyset$, then it is
      isometric to $(\mathbb{R} \times \Sigma, \, t_a t_b - h_{ab})$,
      where $(\Sigma, \, h_{ab})$ is a compact Riemannian manifold and
      $t^a$ is a timelike vector-field in $\mathcal{M}$.  (In
      particular, $(\mathcal{M}, \, g_{ab})$ must be globally
      hyperbolic and static.)

      I discuss the physical meaning of the splitting theorems below.}
    \cite{yau-problems,galloway-horta-reg-lorentz-busemann-fns}
      \item a given globally hyperbolic extension of a spacetime is
    the maximal such extension \cite{ringstrom-cauchy-prob-gr}
  \end{enumerate}
    \item[DEC] \hspace{1em}
  \begin{enumerate}
      \item formation of a closed trapped surface after gravitational
    collapse of arbitrary (\emph{i}.\emph{e}., not necessarily close
    to spherical) matter distribution
    \cite{schoen-yau-exist-bh-condens-matt}
      \item a stationary, asymptotically flat black hole is
    topologically $\mathbb{S}^2$ \footnote{This is also a constituent
      of the proof of the full No Hair theorem, but is important
      enough a result to warrant its own entry in the list; see
      footnote~\ref{fn:no-hair-thm}.  Hawking's original proof was not
      rigorous; in particular, it did not completely rule out a
      toroidal topology.  See \citeN{gannon-topo-slhs-sing-bhs} for a
      rigorous proof of the theorem in electrovac spacetimes, and
      \citeN{galloway-topo-bhs,chrusciel-wald-topol-stat-bhs} for a
      rigorous proof using the NEC{} for otherwise arbitrary
      stress-energy tensors but more stringent constraints on the
      global topology of the spacetime.}  \cite{hawking-bhs-in-gr}
      \item a generalized black hole is topologically $\mathbb{S}^2$
    \footnote{See footnote~\ref{fn:genl-bhs}.}
    \cite{hayward-genl-laws-bhdyns}
      \item constituents of the black hole ``No Hair'' theorems for
    asymptotically flat black holes\footnote{See
      footnote~\ref{fn:no-hair-thm}.}
    \cite{bekenstein-nonexist-baryon-number-stat-bh,carter73,hawking-ellis-lrg-scl-struc-st}
      \item Zeroth Law of black-hole mechanics\footnote{The surface
      gravity is constant on the event horizon of a stationary
      asymptotically flat black hole.}
    \cite{bardeen-carter-hawking73}
      \item Zeroth Law of generalized black-hole
    mechanics\footnote{The total trapping gravity of a generalized
      black hole is bounded from above, and achieves its maximal value
      if and only if the trapping gravity is constant on the trapping
      horizon, which happens when the horizon is stationary.  (See
      footnote~\ref{fn:genl-bhs}.)} \cite{hayward-genl-laws-bhdyns}
      \item every past timelike geodesic in spatially open,
    non-rotating spacetimes with non-zero spatially averaged energy
    densities is incomplete\footnote{This theorem is particularly
      strong: it implies that any singularity-free spacetime
      satisfying the other conditions must have everywhere vanishing
      averaged spatial energies.}
    \cite{senovilla-sing-thm-spatial-avgs,senovilla-new-type-sing-thm}
      \item positivity of ADM energy
    \cite{schoen-yau-pos-mass-thm-ii,witten-pos-energy-thm}
      \item positivity of Bondi energy
    \cite{horowitz-perry-grav-en-cant-neg,ludvigsen-vickers-pos-bondi-mass,schoen-yau-bondi-mass-pos,hayward-grav-enrg-spher-sym}
      \item asymptotic energy-area inequality in the spherically
    symmetric case\footnote{This inequality, first conjectured by
      \citeN{penrose-nkd-sings}, states that if a spacelike
      hypersurface in a spherically symmetric, asymptotically flat
      spacetime contains an outermost marginally trapped sphere of
      radius $R$ (in coordinates respecting the spherical symmetry),
      then the ADM energy $\ge \half R$.  The DEC need hold only on
      the spacelike hypersurface, not in the whole spacetime.}
    \cite{hayward-grav-enrg-spher-sym}
      \item if a covariantly divergence-free $T_{ab}$ vanishes on a
    closed, achronal set, it vanishes in the domain of dependence of
    that set \cite{hawking-cons-matt-gr,hawking-ellis-lrg-scl-struc-st}
      \item standard statements of the initial-value formulation of
    the Einstein field equation with non-trivial $T_{ab}$ is well
    posed (in the sense of Hadamard)
    \cite{hawking-ellis-lrg-scl-struc-st,wald-gr}
      \item natural definition of the center of mass, multipole
    moments and equations of motion for an extended body
    \cite{dixon-dyns-exten-bods-gr-i,dixon-dyns-exten-bods-gr-ii,dixon-defn-multip-moms-exten-bods,dixon-dyns-exten-bods-gr-iii,ehlers-rudolph-dyns-exten-bods-gr,schattner-com-in-gr,schattner-uniq-com-in-gr,ehlers-folklore-rel}
      \item some standard forms of the Cosmic Censorship Hypothesis
    \cite{geroch-horowitz-glob-struc,penrose-sing-time-asym,wald-gr,joshi-cosmic-censor}
  \end{enumerate}
    \item[SDEC] \hspace{1em}
  \begin{enumerate}
      \item geodesic theorem for ``arbitrarily small'' bodies,
    neglecting self-gravitational effects
    \cite{geroch-jang-motion-gr,malament-fnds-gr-ngt,weatherall-econds-geroch-jang}
      \item geodesic theorem for ``arbitrarily small'' bodies,
    including self-gravitational effects
    \cite{ehlers-geroch-eom-small-bods-gr}
  \end{enumerate}
    \item[ANEC] \hspace{1em}
  \begin{enumerate}
      \item focusing theorems for congruences of causal geodesics
    \cite{borde-geo-focus-econds-sings}
      \item formation of singularities after gravitational collapse in
    spatially open spacetimes
    \cite{roman-awec-penrose-sing-thm,senovilla-sing-thms-conseq}
      \item Topological Censorship
    Theorem\footnote{\label{fn:topol-censor}The theorem states: fix an
      asymptotically flat, globally hyperbolic spacetime satisfying
      the ANEC; let $\gamma$ be a causal curve with endpoints on past
      and future null infinity that lies in a simply connected
      neighborhood of null infinity; then every causal curve with
      endpoints on past and future null infinity is smoothly
      deformable to $\gamma$.  Roughly speaking, this theorem says
      that no observer remaining outside a black hole can ever have
      enough time to probe the spatial topology of spacetime:
      isolated, non-trivial topological structure with positive energy
      will collapse into black holes too quickly for light to cross
      it.  In other words, the region outside black holes is
      topologically trivial.}  \cite{friedman-et-topol-censor}
      \item prohibition on traversable wormholes
    \cite{morris-et-wormh-time-mach-wec}
      \item prohibition on spatial topology change
    \cite{borde-topol-chng-gr}
      \item positivity of ADM energy
    \cite{penrose-et-pos-mass-focus-retard-null-geos}
  \end{enumerate}
    \item[AWEC] $\emptyset$
    \item[ASEC] \hspace{1em}
  \begin{enumerate}
      \item cosmological singularities in spatially closed spacetimes
    \cite{tipler-econds-st-sings,senovilla-sing-thms-conseq}\footnote{\label{fn:tipler-asec-constraint}Strictly
      speaking, Tipler's proof requires the ASEC with the additional
      constraint that its characteristic integral can equal 0 for any
      geodesic only if its integrand ($R_{mn} \xi^m \xi^n$) equals 0
      along the entire geodesic.  Senovilla's proof does not require
      these extra assumptions, though it does require the existence of
      a Cauchy surface with vanishing second fundamental-form.}
      \item cosmological singularities in spatially open spacetimes
    \cite{tipler-econds-st-sings,senovilla-sing-thms-conseq}\footnote{Strictly
      speaking, Tipler's proof of this theorem requires the WEC as
      well as the ASEC, and also requires the same further constraint
      on the ASEC as described in
      footnote~\ref{fn:tipler-asec-constraint}.  Senovilla's proof is
      also the same as that described in
      footnote~\ref{fn:tipler-asec-constraint}.}
  \end{enumerate}
\end{description}

There is a striking absentee from the list of consequences: strictly
speaking, the First Law of black-hole mechanics (for asymptotically
flat black holes)---conservation of mass-energy---does not require for
its validity the assumption of any energy condition (unlike the other
three Laws).\footnote{\citeN{hayward-genl-laws-bhdyns} does give a
  proof of what he calls the First Law for generalized black holes
  (footnote~\ref{fn:genl-bhs}), and that does explicitly require the
  NEC{}, but the physical interpretation of Hayward's result is vexed
  (as he himself admits), so I did not list it among the consequences
  of the NEC.  The physical interpretation of that result would be an
  interesting problem to resolve, as it would likely shed light on the
  already vexed problem of understanding energy in general
  relativity.}  The issue is somewhat delicate in the details,
however.  The delicacy arises from the fact that all the most rigorous
and the most physically compelling derivations of the Law I know
\cite{bardeen-carter-hawking73,wald-gao-proc-1st-genl-2nd-charged-rot-bhs}
assume that the surface gravity of the black hole is constant on the
event horizon.  This, of course, is the Zeroth Law of black-hole
mechanics, and all known proofs of the most general form of the Zeroth
Law rely on the DEC.  The qualification ``most general'' is required
because there are weaker forms of the Zeroth Law that require no
energy condition for their proof: any sufficiently regular Killing
horizon must be bifurcate, and the appropriate generalization of
surface gravity for a bifurcate Killing horizon must be constant on
the entire horizon, without the need to impose any energy condition
\cite{kay-wald-uniq-therm-props-stat-bif-khoriz,racz-wald-ext-sts-khoriz,wald-qft-cst,racz-wald-glob-ext-final-stat-bhs,heusler-bh-uniq}.\footnote{Roughly
  speaking, a Killing horizon is sufficiently regular in the relevant
  sense if: it is (locally) bifurcate; or the null geodesic congruence
  constituting it is geodesically complete; or the twist of the null
  geodesic congruence has vanishing exterior derivative; or the domain
  of exterior communication is static; or the domain of exterior
  communication is stationary, axisymmetric, and the 2-surfaces
  orthogonal to the two Killing fields are hypersurface orthogonal.}
This is a weaker form of the Zeroth Law, in so far as it is not known
whether the event horizons of all ``physically reasonable'' black
holes are sufficiently regular in the sense required, though in fact
the event horizons of all known exact black-hole solutions are, and
the condition of sufficient regularity has strong physical
plausibility on its own, at least if one accepts any version of Cosmic
Censorship---it almost necessarily follows that any non-sufficiently
regular horizon will eventuate in a naked singularity.

Whether one considers the First Law a consequence of the DEC,
therefore, depends on whether one thinks it suffices simply to assume
the Zeroth Law in its most general form, whether one thinks one should
include a derivation of the most general form of the Zeroth Law in a
derivation of the First Law, or whether one thinks that the weaker
form of the Zeroth Law, which requires no energy condition, suffices
for the purposes of the First Law.  The delicacy is exacerbated by the
fact that (at least) two conceptually distinct formulations of the
First Law appear in the literature, what (following \citeNP[ch.~6,
\S2]{wald-qft-cst}) I will call the physical-process version and the
equilibrium version.  The former fixes the relations among the changes
in an initially stationary black hole's mass, surface gravity, area,
angular velocity, angular momentum, electric potential and electric
charge when the black hole is perturbed by throwing in an
``infinitesimally small'' bit of matter, after the black hole settles
back down to stationarity.  The latter considers the relation among
all those quantities for two black holes in ``infinitesimally close''
stationary states, or, more precisely, for two ``infinitesimally
close'' black-hole spacetimes.

The roles the assumption of the Zeroth Law plays in the proofs of the
two versions of the First Law differ significantly, moreover, so it is
not clear one could give a single principled answer to the question of
whether or not the First Law is a consequence of the DEC that covered
both versions at once.  For example, in the physical-process version,
but not in the equilibrium version, one must assume that the black
hole settles back down to a stationary state after one throws in the
small bit of matter, and so, \emph{a fortiori}, that the event horizon
is not destroyed when one does so, resulting in a naked singularity.
I know of no rigorous proofs of the stability of an event horizon
under generic small perturbations.  All the most compelling arguments
in favor of a reasonably broad kind of stability I know, however, do
assume constraints on the form of the matter causing the perturbation,
constraints that usually look a lot like energy
conditions.\footnote{See, \emph{e}.\emph{g}.,
  \citeN{press-teukolsky-perturb-rot-bh-ii},
  \citeN{kay-wald-lin-stab-schwarz},
  Carter~\citeyear{carter-bh-equil-solved,carter-bhequil-prob}, and
  \citeN{kokkotas-schmidt-quasi-norm-modes-bhs}.}  Why is there this
problem with understanding the relation of the First Law to the energy
conditions?  The difficulty seems especially surprising in light of
the fact that it is the only one of the Laws that constrains
mass-energy!  Is it, perhaps, that mere conservation doesn't care
whether mass-energy is negative or positive?

As striking as the difficulty in that case is, however, I still find
more striking the number, variety and depth of what are indubitably
consequences that the energy conditions \emph{do} have, especially
without input from the Einstein field equation.  The two most numerous
types of theorems in the list of consequences are those pertaining to
singularities and those to black holes (including horizons),
respectively, at 10 each.  Indeed, it was the epoch-making result of
\citeN{penrose-grav-coll-st-sings} showing that a singularity would
inevitably result from gravitational collapse in an open universe that
first demonstrated the power that the qualitative abstraction of
energy conditions gives in proving far-reaching results of great
physical importance.  I will first discuss some interesting features
of the singularity theorems and the role that energy conditions play
in their proofs, then do the same for theorems about black holes,
positive energy, geodesic theorems and entropy bounds.\footnote{I will
  not discuss the role of energy conditions in ensuring that the
  initial-value formulation of general relativity is well posed, as
  the relation between the two is complex and very little is known
  about it.  That is work for a future project.}  In
\S\ref{sec:violations}, I will then review the violations of the
energy conditions and discuss whether they give us grounds for
doubting the physical relevance of the positive consequences.

The weakest condition, the NEC, already has remarkably strong
consequences.  Among the singularity theorems it supports, to my mind
the most astonishing is the one due to
\citeN{gannon-sings-nonsimp-conn-sts}: in any asymptotically flat
spacetime with a non-simply connected Cauchy surface, a singularity is
bound to form.  Topological complexity by itself, with the only
constraint on metrical structure being the mild one of the NEC,
suffices for the formation of singularities (in the guise of the
incompleteness of a causal geodesic).  The theorem gives one no
information about the singularity, whether it will be a timelike or
null geodesic that is incomplete, or whether it will be associated
with pathology in the curvature, or something that looks like collapse
of a material body, or will be cosmological in character (such as a
Big Bang or Big Crunch), but the simple fact that non-trivial topology
plus the weakest energy condition, irrespective of dynamics, suffices
for geodesic incompleteness already shows the profound power of these
conditions.  It is tempting to relate Gannon's singularity theorem to
Topology Censorship, especially in so far as the latter requires only
the ANEC, which the NEC{} implies.  If one assumes that the
singularity predicted by Gannon's theorem will be hidden behind an
event horizon, then the theorem gives some insight into why
non-trivial spatial topological structure will always (quickly come to
be) hidden inside a black hole.  (See footnote~\ref{fn:topol-censor}.)
It also suggests that, in some rough sense, non-trivial topological
structure may have mass-energy associated with it (perhaps of an
ADM-type).  It would be of some interest to see whether that idea can
be made precise; one possible approach would be to see whether one
could attribute some physically reasonable, non-zero ADM-like mass to
flat, topologically non-trivial spacetimes.  If so, I think this would
give insight into the vexed question of the meaning of ``mass'' and
``energy'' in general relativity.  If such a definition were to be
had, I conjecture that non-trivial topological structure could have
either positive or negative mass-energy, depending on the form of the
structure; otherwise, it would not seem necessary to assume an energy
condition in order to derive the Topological Censorship
Theorem.\footnote{A good place to start might be the investigation of
  asymptotically flat spacetimes with non-trivial second
  Stiefel-Whitney class, as it is known that such spacetimes cannot
  support a global spinor structure \cite{geroch69,geroch70}.  That
  shows already that there is something physically \emph{outr\'e}
  about those spacetimes.}

Another striking feature of the list is that the only important
consequences of the SEC (and the ASEC) are singularity
theorems,\footnote{Although the proposition that a given globally
  hyperbolic extension of a spacetime is the maximal such extension
  depends for its only known proof on the assumption of the SEC, this
  is not really a counter-example to my claim: roughly speaking, the
  proof works by showing that the given globally hyperbolic extension
  cannot be extended (and so is maximal) because to do so would result
  ``immediately'' in singularities, contradicting the assumption of
  extendibility.}  and among them the most physically salient ones,
whereas the DEC, contrarily, is used in only one type of singularity
theorem
(Senovilla~\citeyearNP{senovilla-sing-thm-spatial-avgs,senovilla-new-type-sing-thm}),
and that of a character completely different from the other
singularity theorems.  The singularity theorems following from the SEC
are the most physically salient both because they tend to have the
weakest ancillary assumptions, and because they apply to physically
important situations, both for collapsing bodies and for cosmology.  I
have no compelling explanation for why the SEC should have no
important consequences other than singularity theorems.  Perhaps it
has to do with the fact that the SEC has a relatively clear
geometrical interpretation (convergence of timelike geodesics) that is
manifestly relevant to the formation of singularities, whereas its
physical and effective interpretations are obscure at best.  If so,
then one may want to consider the SEC a case of gerrymandering, the
relativity community simply having posited the weakest formal
condition it could find to prove the results it wants.  This line of
thought becomes especially attractive when one contemplates the many
possible violations of the SEC and even more the strong preponderance
of indirect observational evidence that the SEC has been widely
violated on cosmological scales at many different epochs in the actual
universe, and is likely being violated right now.\footnote{See
  \S\ref{sec:violations} for discussion, and
  \citeN{curiel-econds-cosmo} for a more extensive and thorough
  analysis.}  The result of \citeN{ansoldi-spher-bhs-reg-cntr},
however, that black holes with singularity-free interiors necessarily
violate the SEC, may push one towards the opposite view, in so far as
it comes close to making the SEC both necessary and sufficient for the
occurrence of certain types of singularities.  (The construction of
singularity-free FLRW spacetimes violating the SEC, in
\citeNP{bekenstein-non-sing-gr-cosmo}, buttresses this line of
thought; I discuss this further below.)

I have no explanation for why the DEC should be used in almost no
singularity theorems, except for the simple observation that the only
real addition the DEC makes to the NEC{} and the WEC, that
energy-momentum flux be causal, has no obvious connection to the
convergence of geodesics.  The one type of singularity theorem
(Senovilla~\citeyearNP{senovilla-sing-thm-spatial-avgs,senovilla-new-type-sing-thm})
it is used in, moreover, is the only one to make substantive, explicit
assumptions (over and above the energy conditions themselves) about
the distribution of stress-energy, in this case in the demand for
non-zero averaged spatial energy density.  Perhaps that is why the DEC
comes into play in this theorem, though I have no real insight into
how or why the DEC may bear on averaged spatial energy density and its
relation to the convergence of geodesic congruences.

The Lorentzian splitting theorems may be thought of as rigidity
theorems for singularity theorems invoking the SEC, for the splitting
theorems show that, under certain other assumptions, there will be no
singularities only when the spacetime is static and globally
hyperbolic.  Static and globally hyperbolic spacetimes, however, are
``of measure zero'' in the space of all spacetimes, and so being free
of singularities is, under the ancillary conditions, unstable under
arbitrarily small perturbations.\footnote{One should bear in mind that
  this argument is hand-waving at best.  First, there is no known
  natural measure on the space of spacetimes; second, even if there
  were, being a measure on an infinite-dimensional space, it is
  possible that open sets (in some natural topology, of which there is
  also not one known) would have measure zero.  (It is a well known
  theorem that there is no Lebesgue measure on an infinite-dimensional
  Banach space; thus measure and topology tend to come apart.)  In
  that case, in a natural sense ``arbitrarily small'' perturbations of
  a static, globally hyperbolic spacetime could in fact yield another
  static, globally hyperbolic spacetime.  This problem is not unique
  to this argument but plagues all hand-waving arguments invoking
  ``measure zero'' sets in the space of all spacetimes, which are a
  dime a dozen, especially in the cosmology literature.}  Thus, they
go some way towards proving the conjecture of
\citeN{geroch-sings-closed-uni} that essentially all spatially closed
spacetimes either have singularities or do not satisfy the
SEC.\footnote{If this conjecture were to be precisely formulated and
  proven, perhaps one could view it as providing something like an
  \emph{a posteriori} partial physical interpretation of the SEC.}

As a group, the singularity theorems are perhaps the most striking
example of the importance of ascertaining the status and nature of the
energy conditions, because all the assumptions used in proving
essentially all of them have strong observational or theoretical
support \emph{except} the energy conditions, as
\citeN{sciama-bhs-thermo} emphasized even before there were serious
observational grounds for doubting any of the energy conditions.  This
raises the question of the necessity of the energy conditions for the
singularity theorems.  That some of the impressionist energy
conditions can be used to prove essentially identical theorems already
shows that satisfaction of the pointilliste conditions is not
necessary for validity of at least some of the theorems.  The original
singularity theorem, the demonstration by
\citeN{penrose-grav-coll-st-sings} that singularities should form
after gravitational collapse in spatially open universes, holds under
the weaker assumption of the ANEC
\cite{roman-awec-penrose-sing-thm,senovilla-sing-thms-conseq}.
Likewise, the existence of cosmological (\emph{i}.\emph{e}.,
non-collapse) singularities in both spatially open and closed
universes can be shown under the assumption of the ASEC
\cite{tipler-econds-st-sings,senovilla-sing-thms-conseq}, without the
full SEC.  So far as I know, there is no proof that gravitational
collapse will lead to singularities in the case of spatially closed
spacetimes under the weaker assumption of an impressionist energy
condition.  I conjecture that there are such theorems; it would be of
some interest to formulate and prove one or to construct a
counter-example.

With the possible exception of the First Law of black-hole mechanics
(for asymptotically flat black holes), every fundamental result about
black holes requires an energy condition for its proof, with the
majority relying either on the NEC{} or the DEC.  Roughly speaking,
the results pertaining to black holes fall into three categories:
those constraining the topological and Killing structure of horizons;
those constraining the kinds of property black holes can possess; and
those contraining the relations among the horizon and the properties.
Almost all of the first category invoke the NEC{} for their proof.
One can perhaps see why the NEC{} is relevant for the results about
the topological and Killing stucture of horizons associated with
asymptotically flat black holes: such a black hole is defined as an
event horizon, which is the boundary of the causal past of future
null-infinity, and the boundary of the causal past or future of any
closed set is a null surface, \emph{i}.\emph{e}., is generated by null
geodesics and so may be thought of as a null geodesic congruence.  The
proofs of many of those results, moreover, tend to have the same
structure: very broadly speaking, one assumes the result is not true
and then derives a contradiction with the fact that null geodesic
congruences, by dint of the NEC, must be convergent (or at least not
divergent).  This suggests that the NEC{} is necessary for these
theorems, a suspicion strengthened by the facts that, first, there is
no weaker energy condition that one could attempt to replace it with
(except perhaps the FEC, if it turns out to be viable---see
\S\ref{sec:very-recent-work}), and, second, no such results are known
to follow from any of the impressionist energy conditions.  Again, it
would be of interest to see whether the impressionist energy
conditions could be used to prove theorems about the topological and
Killing structure of black-hole horizons, or else to construct
counter-examples to the results in spacetimes in which the
impressionist but not the pointilliste conditions hold.  The NEC{} is
also used to prove many results about the kinds of properties required
to characterize black holes (the constituents of the ``No Hair''
Theorems), \emph{viz}., that stationary black holes can be entirely
characterized by three parameters, mass, angular momentum and electric
charge.  I have no physically compelling story to tell about why the
NEC{} relates intimately to these kinds of result.  Again, the lack of
such results depending on impressionist conditions suggests that the
pointilliste conditions are necessary, and, again, it is would be of
some interest either to prove analogous results using the
impressionist conditions or to find counter-examples.

Every consequence of the DEC pertaining to black holes is of the kind
that constrains topological or Killing structure of the horizons.
There is, however, no common thread to the role the DEC plays in the
proofs of the various results about black holes it is assumed for,
analogous to the way that the NEC{} plays essentially the same role in
the proofs of many of its consequences.  It is thus difficult even to
hazard a guess about the necessity of the DEC for these consequences.
It would be of great interest to work through the various results to
see whether counter-examples to them satisfying or violating the DEC
could be found, or whether proofs using weaker energy conditions can
be found.  That there is no impressionist analogue to the DEC may
suggest that the DEC is necessary for these results.

Roughly speaking, the idea of the Cosmic Censorship Hypothesis is that
``naked singularities'' should not be allowed to occur in nature,
where, continuing in the same rough vein, a naked singularity is one
that is visible from future null infinity.  Now, the relation of the
energy conditions to the status of the Cosmic Censorship Hypothesis is
complicated, first and foremost, by the fact that there are a
multitude of different formulations of the Hypothesis (thus calling
into question the common practice of honoring the thing with the
capitalization of its name).  Because the presence of naked
singularities would seem to herald a spectacular breakdown in
predictability and even determinism associated with dynamical
evolution in general relativity (such as it is),\footnote{See,
  \emph{e}.\emph{g}., \citeN{earman95} for a thorough discussion, and
  \citeN{curiel-sing} for arguments arriving at somewhat contrary
  conclusions.}  many attempts to make the Hypothesis precise focus on
the initial-value formulation of general relativity.  The most common
formulations invoke either the DEC or the WEC
\cite{joshi-cosmic-censor} as a constraint on the matter fields
permissible for the initial-value formulation of general relativity.
As initially plausible as are such attempts at formulating a precise
version of the Hypothesis that would admit of rigorous proof, there
are in fact cases where satisfaction of an energy condition actually
seems to aid the development of a naked singularity after
gravitational collapse, \emph{e}.\emph{g}., the WEC in the case of the
self-similar collapse of a body of perfect fluid
\cite{joshi-cosmic-censor}.  In such cases, one can show that the
focusing effects the energy condition induces in geodesic congruences
actually contributes directly to the lack of an event horizon.  It is
thus parlous to attempt to draw any concrete conclusions regarding the
relation of the energy conditions to the Cosmic Censorship Hypothesis
in our current state of knowledge.

With regard to results about positivity of global mass, because the
NEC{} does not require the convergence of timelike geodesics (as I
discussed in \S\ref{sec:point-energy-cond}), and so does not entail
that ``gravity be attractive'' for bodies traversing such curves, it
is particularly striking that
\citeN{penrose-light-rays-sl-inf-pos-mass-thm} and
\citeN{ashtekar-penrose-mass-pos-focus-struc-sl-inf} were able to
prove positivity of ADM mass using only it, and that
\citeN{penrose-et-pos-mass-focus-retard-null-geos} were able to prove
it using the even weaker ANEC, and not the significantly stronger DEC,
as all other known proofs require.  All known proofs of the positivity
of the Bondi mass do require the DEC, which is perhaps not surprising,
in light of the fact that the Bondi energy essentially tracks
mass-energy radiated away along null curves to future null infinity.
If the DEC were to fail, then it seems plausible that the Bondi energy
could become negative, if negative mass-energy radiated to null
infinity.  It would be of some interest to try to find a spacetime
model with negative Bondi mass in which the DEC is not violated.
Perhaps matter fields with ``superluminal acoustic modes'' that still
satisfied the DEC (\S\ref{sec:point-energy-cond}) might provide such
examples.

The most precise, rigorous and strongest geodesic theorems
\cite{geroch-jang-motion-gr,ehlers-geroch-eom-small-bods-gr} both
assume the SDEC.\footnote{The statement of the theorems in each of
  those papers in fact uses the DEC, but an examination of the proof
  shows that they both actually use the SDEC, in both cases in order
  to ensure that a constructed scalar quantity that can be thought of
  as the mass of an ``arbitrarily small'' body is strictly greater
  than zero.}  Under the assumptions used to prove the theorem of
\citeN{geroch-jang-motion-gr}, \citeN{malament-geod-princ-gr} showed
that the SDEC is necessary for the body to follow a geodesic, and not
just any timelike curve.  \citeN{weatherall-econds-geroch-jang}
strengthened the result by showing that the SDEC is necessary for the
geodesic to be timelike, not spacelike.  He showed as well that the
SDEC is not strong enough to ensure that the curve not be null: there
is a spacetime with a null geodesic satisfying all the conditions of
the Geroch-Jang Theorem.  It is perhaps important that the example
\citeN{weatherall-econds-geroch-jang} produces to show that a null
curve can satisfy all of the theorem's conditions relies on a
stress-energy tensor not of Hawking-Ellis type~\textsc{i}.  Since
stress-energy tensors not of type~\textsc{i} are generally considered
``unphysical'', it would be of interest to determine whether there are
counter-examples to the \citeN{geroch-jang-motion-gr} and
\citeN{ehlers-geroch-eom-small-bods-gr} theorems that rely on
stress-energy tensors of type~\textsc{i}.  Because of the character of
the proofs of the theorems and of the counter-examples that
\citeN{weatherall-econds-geroch-jang} produces, I conjecture that
there are no such counter-examples, and thus that violations of the
theorems require non-standard stress-energy tensors.\footnote{I have
  not had the opportunity to work through the arguments of
  Dixon~\citeyear{dixon-dyns-exten-bods-gr-i,dixon-dyns-exten-bods-gr-ii,dixon-defn-multip-moms-exten-bods,dixon-dyns-exten-bods-gr-iii},
  \citeN{ehlers-rudolph-dyns-exten-bods-gr} and
  Schattner~\citeyear{schattner-com-in-gr,schattner-uniq-com-in-gr} to
  determine whether their results on the definability of the center of
  mass of an extended body and the formulation of equations of motion
  for that center of mass in fact rely on the SDEC rather than, as
  they explicitly assumed, the DEC.  Because of the intimate
  connection of these relations with the geodesic theorems, this would
  be of some interest to determine.}

Whether or not my conjecture is correct, I think the necessity of the
strongest energy condition for the validity of the theorems poses a
problem for many attempts to analyze and clarify the conceptual
foundations of general relativity.  Many attempts to provide
interpretations of the formalism of general relativity, for instance,
place fundamental weight on the so-called Geodesic Principle, that
``small bodies'', when acted on by no external forces, traverse
timelike geodesics.  The ``fact'' that the Geodesic Principle is a
consequence of the Einstein field equation is often cited as
justification for the validity of the Principle (\emph{e}.\emph{g}.,
\citeNP{brown-phys-rel}).  The work of \citeN{malament-geod-princ-gr}
and \citeN{weatherall-econds-geroch-jang}, however, show that, at
best, such approaches to the foundations of general relativity must be
more subtle where the Geodesic Principle is concerned, and, at worst,
that the Principle may in fact not be suitable at all for playing a
fundamental role in giving an interpretation of the theory.

With regard to entropy bounds such as that of
Bousso~\citeyear{bousso-cov-ent-conj,bousso-holo-gen-st}, if in fact
the NEC{} or DEC were necessary for their validity, this could spell
serious trouble for many programs in quantum gravity, or at least for
the ways that research in such programs are currently being carried
out, in so far as many programs place enormous motivational,
argumentative and interpretational weight on such entropy bounds, and
we already know that essentially all energy conditions are
promiscuously violated when quantum effects are taken into
account.\footnote{See \citeN{curiel-econds-qftcst} for more detailed
  discussion of all these issues.}

\subsection{Violations}
\label{sec:violations}

\begin{description}
    \item[NEC] \hspace{1em}
  \begin{enumerate}
      \item conformally coupled massless and massive scalar fields
    \textbf{[possibly]}
    \cite{visser-barcelo-econds-cosmo-implic,barcelo-visser-twilight-econds}
      \item generically non-minimally coupled massless and massive
    scalar fields \textbf{[possibly]}
    \cite{flanagan-wald-backreact-null-energy-semiclass,visser-barcelo-econds-cosmo-implic,barcelo-visser-twilight-econds,dubovsky-et-nec-superlum-prop}
      \item ``big bang'' and ``big crunch''
    singularities\footnote{\label{fn:big-bang}A big bang or a big
      crunch is a singularity in a standard cosmological model where
      the expansion factor $a(t) \rightarrow 0$ in a finite period of
      time to the past or future, respectively.  See,
      \emph{e}.\emph{g}., \citeN{weinberg72} or \citeN{wald-gr}.  In
      the specific context of FLRW spacetimes, this condition implies
      that a singularity is ``strong'' in the sense of
      \citeN{tipler-sings-conf-flat-sts}.}  \textbf{[possibly]}
    \cite{cattoen-visser-necsuff-cosmo-events,cattoen-visser-cosmodyns-econds-bounds}
      \item ``big rip'' singularities\footnote{A big rip is a
      singularity in a standard cosmological model where the expansion
      factor $a(t) \rightarrow \infty$ in a finite period of time.
      If, as is currently believed, the universe is expanding at an
      accelerated rate, and it continues to do so, it is possible that
      such a big rip will occur.  See, \emph{e}.\emph{g}.,
      \citeN{caldwell-phantom-menace},
      \citeN{caldwell-et-phantom-energy-cosm-doom} and
      \citeN{chimento-lazkoz-big-rip-sings}.}  \textbf{[necessarily]}
    \cite{cattoen-visser-necsuff-cosmo-events,cattoen-visser-cosmodyns-econds-bounds}
      \item sudden future
    singularities\footnote{\label{fn:sudden-sings}These are
      singularities in standard cosmological models in which the
      pressure of the effective fluid or some higher derivative of the
      expansion factor $a(t)$ diverges, even though the energy density
      and curvature remain well behaved.  They are very strange, not
      least because they do not necessarily lead to
      curve-incompleteness of any kind.  See
      \citeN{curiel-anal-sings-revis} for further discussion.}
    \textbf{[possibly]}
    \cite{barrow-more-genl-sudd-sings,barrow-sudd-fut-sings,cattoen-visser-necsuff-cosmo-events,cattoen-visser-cosmodyns-econds-bounds}
      \item naked singularities \textbf{[possibly]}
    \cite{penrose-sing-time-asym,barcelo-visser-travers-worms-massl-conform-sfs,joshi-cosmic-censor}
      \item closed timelike curves \textbf{[possibly]}
    \cite{visser-lorentz-worms}
      \item Tolman wormholes and Einstein-Rosen bridges
    \textbf{[necessarily]}
    \cite{barcelo-visser-travers-worms-massl-conform-sfs}
      \item any fluid with a barotropic index $w < -1$ \footnote{See
      footnote~\ref{fn:w-index}.}  (such as those postulated in
    so-called phantom cosmologies) \textbf{[necessarily]}
    \cite{visser-lorentz-worms,dabrowski-denkiewicz-baro-w-sings-cosmo}
      \item ``hyper-fast'' travel\footnote{See
      footnote~\ref{fn:hyperfast}.}  \textbf{[possibly]}
    \cite{visser-et-superlum-cens}
  \end{enumerate}
    \item[WEC] \hspace{1em}
  \begin{enumerate}
      \item naked singularities \textbf{[possibly]}
    \cite{ford-roman-cosmic-flashing}
      \item closed timelike curves \textbf{[possibly]}
    \cite{visser-lorentz-worms}
      \item physically traversable wormholes \textbf{[necessarily]}
    \cite{morris-thorne-worms-st-interstell-trvl,visser-travers-wormhs-exs,visser-travers-wormhs-mod-schwrz}
      \item cosmological steady-state theories of
    \citeN{bondi-gold-steady-state} and
    \citeN{hoyle-new-model-expand-uni}\footnote{See also
      \citeN{pirani-energy-creat-matter-rel-cosmo},
      \citeN{hoyle-narlikar-new-theory-grav}, and \citeN[\S4.3,
      pp.~90--91; \S5.2, p.~126]{hawking-ellis-lrg-scl-struc-st}.}
    \textbf{[necessarily]}
      \item classical Dirac fields \textbf{[possibly]}
    \cite{wald-gr}
      \item a positive cosmological constant (\emph{e}.\emph{g}.,
    anti-de{} Sitter Space)\footnote{It should be kept in mind that
      the physical consequences of a ``positive'' versus a
      ``negative'' cosmological constant in this context depend on
      one's conventions for writing the Einstein field equation and on
      one's conventions for the metric signature.  With the
      conventions I am using, a negative value of $\Lambda$ itself
      leads to negative momentum flux in spacelike directions, and
      that is the condition that leads to accelerated expansion on the
      cosmological scale, as actually observed, and so the theoretical
      need for ``dark energy''.}  \textbf{[necessarily]}
    \cite{hawking-ellis-lrg-scl-struc-st,visser-lorentz-worms}
      \item future-eternal inflationary cosmologies
    \textbf{[possibly]} \cite{borde-vilenkin-viol-wec-inflat}
      \item ``hyper-fast'' travel\footnote{See
      footnote~\ref{fn:hyperfast}.} \textbf{[necessarily]}
    \cite{alcubierre-warp-drive,krasnikov-hyperfast-trvl-gr,olum-superlum-trvl-req-neg-energ}
  \end{enumerate}
    \item[SEC] \hspace{1em}
  \begin{enumerate}
      \item ``big bang'' and ``big crunch'' singularities\footnote{See
      footnote~\ref{fn:big-bang}.}  \textbf{[possibly]}
    \cite{cattoen-visser-necsuff-cosmo-events,cattoen-visser-cosmodyns-econds-bounds}
      \item sudden future singularities\footnote{See
      footnote~\ref{fn:sudden-sings}.}  \textbf{[possibly]}
    \cite{barrow-more-genl-sudd-sings,barrow-sudd-fut-sings,cattoen-visser-necsuff-cosmo-events,cattoen-visser-cosmodyns-econds-bounds}
      \item cosmological ``bounces''\footnote{A bounce, in the context
      of a standard cosmological model, is a local minimum of the
      expansion factor $a(t)$.  See, \emph{e}.\emph{g}.,
      \citeN{bekenstein-non-sing-gr-cosmo} and
      \citeN{molina-visser-conds-frw-bounce}.}  \textbf{[necessarily]}
    \cite{cattoen-visser-necsuff-cosmo-events,cattoen-visser-cosmodyns-econds-bounds}
      \item just before or just after a cosmological
    ``inflexion''\footnote{An inflexion, in the context of a standard
      cosmological model, is a saddle-point of the expansion factor
      $a(t)$.  See, \emph{e}.\emph{g}., \citeN{sahni-et-loiter-uni}
      and \citeN{sahni-shtanov-uni-loiter-high-red-shifts}.}
    \textbf{[possibly]}
    \cite{cattoen-visser-necsuff-cosmo-events,cattoen-visser-cosmodyns-econds-bounds}
      \item spatially closed, expanding, singularity-free spacetimes
    \textbf{[necessarily]} \cite{senovilla-sing-thms-conseq}
      \item cosmological inflation \textbf{[necessarily]}
    \cite{visser-lorentz-worms}
      \item a negative cosmological constant, as in de{} Sitter
    spacetime, and the ``dark energy'' postulated to drive the
    observed accelerated expansion of the universe
    \textbf{[necessarily]}
    \cite{hawking-ellis-lrg-scl-struc-st,caldwell-phantom-menace,caldwell-et-phantom-energy-cosm-doom,dabrowski-et-phantom-cosmo}
      \item asymptotically flat black holes with regular
    (non-singular) interiors \textbf{[necessarily]}
    \cite{ansoldi-spher-bhs-reg-cntr}
      \item closed timelike curves \textbf{[possibly]}
    \cite{visser-lorentz-worms}
      \item physically traversable wormholes \textbf{[necessarily]}
    \cite{molina-visser-conds-frw-bounce,hochberg-et-tolman-worms-viol-sec}
      \item minimally coupled massless and massive scalar fields
    \textbf{[possibly]}
    \cite{visser-barcelo-econds-cosmo-implic,barcelo-visser-twilight-econds}
      \item massive Klein-Gordon fields \textbf{[possibly]}
    \cite{visser-lorentz-worms}
      \item typical gauge theories with spontaneously broken
    symmetries \textbf{[possibly]} \cite{tipler-econds-st-sings}
      \item conformal scalar fields coupled with dust
    \textbf{[possibly]} \cite{bekenstein-non-sing-gr-cosmo}
      \item ``hyper-fast'' travel\footnote{See
      footnote~\ref{fn:hyperfast}.} \textbf{[necessarily]}
    \cite{krasnikov-hyperfast-trvl-gr,olum-superlum-trvl-req-neg-energ,alcubierre-warp-drive}
  \end{enumerate}
    \item[DEC] \hspace{1em}
  \begin{enumerate}
      \item ``big bang'' and ``big crunch'' singularities\footnote{See
      footnote~\ref{fn:big-bang}.}  \textbf{[possibly]}
    \cite{cattoen-visser-necsuff-cosmo-events,cattoen-visser-cosmodyns-econds-bounds}
      \item sudden future singularities\footnote{See
      footnote~\ref{fn:sudden-sings}.}  \textbf{[possibly]}
    \cite{barrow-more-genl-sudd-sings,barrow-sudd-fut-sings,cattoen-visser-necsuff-cosmo-events,cattoen-visser-cosmodyns-econds-bounds}
      \item classical Dirac fields \textbf{[necessarily]}
    \cite{penrose-rindler-spinors-st-1}
  \end{enumerate}
    \item[ANEC] \hspace{1em}
  \begin{enumerate}
      \item massless conformally coupled scalar
    fields\footnote{\citeN{urban-olum-anec-viol-conf-flat-st} also
      show that AANEC can be violated by conformally coupled scalar
      fields in conformally flat spacetimes, such as the standard FLRW
      cosmological models.}  \textbf{[possibly]}
    \cite{visser-barcelo-econds-cosmo-implic,barcelo-visser-twilight-econds}
      \item massless and massive non-minimally coupled scalar fields
    \textbf{[possibly]}
    \cite{flanagan-wald-backreact-null-energy-semiclass,dubovsky-et-nec-superlum-prop}
      \item closed timelike curves \textbf{[possibly]}
    \cite{visser-lorentz-worms}
      \item traversable wormholes \cite{morris-et-wormh-time-mach-wec}
    \textbf{[possibly]}
  \end{enumerate}
    \item[AWEC] \hspace{1em}
  \begin{enumerate}
      \item cosmological steady-state theories of
    \citeN{bondi-gold-steady-state} and
    \citeN{hoyle-new-model-expand-uni} \textbf{[necessarily]} (my
    calculation)
      \item a positive cosmological constant (\emph{e}.\emph{g}.,
    anti-de{} Sitter Space) \textbf{[necessarily]} (my calculation)
      \item classical Dirac fields \textbf{[possibly]} (my
    calculation)
      \item closed timelike curves \textbf{[possibly]}
    \cite{visser-lorentz-worms}
      \item physically traversable wormholes \textbf{[possibly]} (my
    calculation)
      \item ``hyper-fast'' travel\footnote{See
      footnote~\ref{fn:hyperfast}.}  \textbf{[possibly]} (my
    calculation)
  \end{enumerate}
    \item[ASEC] \hspace{1em}
  \begin{enumerate}
      \item a negative cosmological constant, as in de{} Sitter
    spacetime, and the ``dark energy'' postulated to drive the
    observed accelerated expansion of the universe
    \textbf{[necessarily]} (my calculation)
      \item cosmological inflation \textbf{[possibly]} (my
    calculation)
      \item massive Klein-Gordon fields \textbf{[possibly]} (my
    calculation)
      \item typical gauge theories with spontaneously broken
    symmetries \textbf{[possibly]} (my calculation)
      \item conformal scalar fields coupled with dust
    \textbf{[possibly]} (my calculation)
  \end{enumerate}
\end{description}

The most compelling empirical evidence for violations of energy
conditions comes from cosmology.  For instance, strongly substantiated
cosmographic arguments comparing best estimates for the age of the
oldest stars to the epoch of galaxy formation show that the SEC must
have been violated in the relatively recent cosmological past
(redshift $z < 7$)
\cite{visser-econds-epoch-galax-form,visser-gr-econds-hubble-expan,visser-cosmography}.
Visser's arguments, especially as presented in the 1997 papers, are an
especially striking example of the power of the energy conditions:
years before there was any hard observational evidence for the
acceleration of the current expansion of the universe, and so hard,
direct support for the existence of a negative cosmological constant,
Visser predicted on purely theoretical grounds that the most likely
culprit for violation of SEC in the recent cosmological past must be a
negative cosmological constant.  In fact, if the current consensus
that the expansion of the universe is accelerating is correct, and so
some form of ``dark energy'' exists, then we know that the SEC is
currently being violated on cosmological scales, entirely
independently of any assumptions about the nature of the fields
entering into the stress-energy tensor or cosmological constant
\cite{visser-barcelo-econds-cosmo-implic,barcelo-visser-twilight-econds,visser-cosmography,cattoen-visser-necsuff-cosmo-events,cattoen-visser-cosmo-miles-econds,cattoen-visser-cosmodyns-econds-bounds}.
Finally, if any model of inflationary cosmology is correct, then we
know that the SEC was necessarily violated at least during that period
and, depending on the particulars of the model, possibly the ASEC as
well.  One glimmer of hope among the gloom, however, is that the
presence of a negative cosmological constant does not yield violations
of the NEC, so no matter how exotic so-called dark energy is, and
whatever fundamental mechanism may underlie it, at the classical level
at least it will still satisfy that condition.

Far and away the simplest theoretical mechanisms presently known for
yielding violations of energy conditions, and in many ways the most
plausible, come from models including scalar fields.  Indeed, using
classical scalar fields alone, without even having to resort to
quantum weirdness, it is relatively easy to engineer violations of
even the weakest conditions, the NEC{} and the ANEC, as the list of
violations shows.  We do not yet have indubitable evidence for the
existence of a fundamental scalar field in nature.  (The recently
discovered Higgs field is without question phenomenologically a scalar
field, but the jury is still out on whether or not it is a composite,
bound state of underlying non-scalar entities.)  The importance of
scalar fields in fundamental theoretical physics, however, is
indubitable.\footnote{It would be an interesting project to try to
  determine why theoretical physicists are firmly wedded to scalar
  fields as fundamental constituents of reality in the face of an
  almost complete lack of evidence for them, and whether their reasons
  for the marriage are really sound.}  For many theoretical and
pragmatic reasons, the so-called inflaton field that drives
cosmological inflation is most commonly modeled as a classical scalar
field, and cosmological inflation necessarily violates SEC and,
depending on particulars of the model, possibly ASEC.  Many meson
fields in the Standard Model (pions, kaons and many other mesons,
including their ``charmed'', ``truth'' and ``beauty'' correlates),
moreover, are modeled to an extraordinarily high degree of accuracy as
scalar fields, even though we believe they in fact consist of bound
states of quark-antiquark pairs.  It is also widely believed that the
so-called ``strong CP problem'', the fact that no CP-violation in
strong nuclear interactions has ever been observed, is best solved by
the postulation of a scalar field called the axion
\cite{peccei-quinn-cp-cons-pseuodparts}, though to the best of my
knowledge it is not known whether any classical models of the axion
violate any of the energy conditions (any more than those of other
quantum fields do, at any rate).

Now, violations of the NEC{} are disturbing for at least two important
reasons.  First and perhaps foremost, they imply violations of all
other pointilliste energy conditions.  Second, they already would seem
to allow not only violations of the ordinary Second Law of
thermodynamics
\cite{ford-qcoher-eff-2nd-law,davies-ottewill-detec-neg-enrgy-4d-exs},
but of the Generalized Second Law as well: send lots of negative
energy (with positive entropy) through the event horizon into a black
hole, and \emph{voil\`a}!---the area of the black hole shrinks, even
though arbitrary amounts of entropy have disappeared from outside the
event horizon.  Perhaps the most troubling violation of the NEC{} from
the above list is the case of a conformally coupled scalar field,
given the naturalness of ``conformal coupling'' for scalar fields in
quantum field theory
\cite{visser-barcelo-econds-cosmo-implic,barcelo-visser-twilight-econds},
which is why in the list of violations I singled it out from the class
of generically non-minimally coupled scalar fields.

The particular example of a massive conformal scalar field coupled
with dust given by \citeN{bekenstein-non-sing-gr-cosmo} in an example
of how to construct a nonsingular FLRW model, exploiting the fact that
the system can be made to violate the SEC, has interesting possible
physical significance, which is why I singled it out in the list of
systems for which energy conditions can fail: the pions that mediate
the strong nuclear force can to a very high degree of approximation be
represented by just such scalar fields.  Thus, Bekenstein argues,
nuclear matter in the very early, dense stages of the actual universe
may not have satisfied the SEC, which may suggest that the initial
singularity in standard Big-Bang models may be avoidable.  This may
give reason to doubt the stability of at least some of the singularity
theorems in regimes where the energy conditions fail.  Because the SEC
would have been necessarily violated during an epoch of inflationary
expansion, moreover, and because inflationary theories have such
strong support among many cosmologists, such doubts should perhaps
cause further concern for advocates of an initial Big Bang
singularity.  In light of the fact that the strongest theorems for big
bang singularities rely on the SEC, and the Lorentzian Splitting
Theorems come close to showing that the SEC is necessary for those
theorems, I think it becomes quite reasonable to question the current
confidence in the so-called Standard Model of cosmology, which rests
on the idea that the universe ``started with'' a big bang.  That,
moreover, both a cosmological ``bounce'' and a Tolman wormhole
(perhaps the two most natural possible replacements for an initial big
bang singularity) require violation \emph{only} of the SEC
\cite{hochberg-et-tolman-worms-viol-sec,molina-visser-conds-frw-bounce},
not any of the other energy conditions, only exacerbates the problem.

\citeN{tipler-econds-st-sings}, in a line of argument intended to
mitigate such doubts, has pointed out an amusing poignancy in the role
that homogeneity (high symmetry) plays in Bekenstein's construction of
non-singular FLRW spacetimes that violate the SEC.  It follows from a
theorem Tipler proves that, if a black hole (marginally trapped
surface) develops in one of Bekenstein's spacetimes, then, because
they do satisfy the WEC, a singularity would necessarily develop.  Of
course, a marginally trapped surface would form only if there were
deviations from homogeneity.  We would expect, however, on physical
grounds, that even slight deviations from homogeneity could lead to
the development of marginally trapped surfaces.  Thus, it is only the
strict symmetry of the Bekenstein models that precludes singularities.
This, of course, turns the standard (mistaken)
pre-\citeN{penrose-grav-coll-st-sings} argument on its head: that the
singularities of the FLRW, Schwarzschild, and
\citeN{oppenheimer-snyder-cont-grav-contract} spacetimes were simply
an artifact of their unrealistic perfect symmetry.  In the case of
Bekenstein's spacetimes, it is only their unrealistic perfect
symmetries that \emph{precludes} singularities.  Theorem 1 of
\citeN{tipler-econds-st-sings}, moreover, gives him even stronger
grounds for thinking that violations of SEC will not necessarily block
formation of singularities, at least for closed universes, so long as
the period and extent of the failure is limited with respect to its
satisfaction in the rest of spacetime, \emph{i}.\emph{e}., so long as
the ASEC holds.

The theorems predicting big bang and big crunch singularities face one
more problem peculiar to them alone: all such theorems invoke energy
conditions of various kinds, mostly the SEC, and yet one can show
that, depending on the characteristics of a given big bang or big
crunch singularity, the presence of the singularity itself implies a
violation of the relevant energy condition.  Roughly speaking, whether
a big bang or big crunch implies a violation of a given energy
condition depends on how ``violent'' the singularity is, which idea
can be made precise by analysis of the nature of the matter fields
present (\emph{e}.\emph{g}., the value of the barotropic index of the
ambient homogeneous cosmological fluid), or by the behavior of
geodesic congruences in the immediate neighborhood of the singularity
(\emph{e}.\emph{g}., whether such singularities are strong in the
sense of \citeNP{tipler-econds-st-sings}, and, if so, how quickly they
squeeze spatial volumes to zero).  What is one to say in such cases?
Clearly, the known theorems do not apply to such singularities, but
also clearly the exact spacetimes in which such singularities occur
have been shown to exist.  The only safe conclusion seems to be that,
at least in the case of these kinds of singularity, violations of
salient energy conditions need not preclude their existence.  But then
one must question the importance of the theorems themselves,
especially in light of the growing body of observational evidence
that, if there is a big bang or big crunch, it may well be of a type
that violates energy conditions.

What about the remainder of the singularity theorems?  Should any of
the violations drive us to doubt their validity or physical relevancy?
In order to try to answer this question with some generality, it will
be useful to draw two distinctions, the first between types of
violations, and the second between types of theorems.\footnote{I do
  not think the classifications I sketch here are of relevance beyond
  the context of such discussions as this.  I certainly do not think
  they capture anything of fundamental significance about the nature
  of violations of energy conditions or about singularities.}  First,
roughly speaking, the violations fall into one of two classes, being
associated with a type of physical system (\emph{e}.\emph{g}.,
conformally coupled scalar field, classical Dirac field) or with a
type of ``event'' (very loosely construed, \emph{e}.\emph{g}.,
traversable wormhole, closed timelike curve, or big rip singularity).
Generally speaking, for the latter, the regions where the energy
conditions are violated can be ``localized'' to a neighborhood of the
``event''.  The scare-quotes are to remind us of the fact that some
such events---\emph{e}.\emph{g}., many types of singularities---are
not localizable in any reasonable sense of the term.\footnote{See
  \citeN{curiel-sing} for discussion.}  The qualification ``generally
speaking'' hedges against cases such as the traversable wormholes of
\citeN{visser-travers-wormhs-exs}, for which travelers moving through
the wormholes never experience a violation of any energy condition.
Generally speaking, for violations of the former class (\emph{viz}.,
associated with a type of physical system), one cannot ``localize''
the regions of violation in any way, unless one can localize the
system itself, or at least those spacetime regions in which the system
is known to violate the energy conditions and one can also determine
that the system violates them nowhere else.

As for the singularity theorems, they also fall roughly into two
classes, which for lack of better terms I will refer to as pinpointing
and not.  Roughly speaking, pinpointing theorems, as the name
suggests, in certain ways allow one to say where in spacetime the
singularities occur, and so in a sense one can ``localize'' the
singularities.\footnote{Again, see \citeN{curiel-sing} for discussion
  of why the scare-quotes are called for.}  Such theorems demonstrate
the existence of singularities associated with closed, trapped
surfaces (for singularities contained in asymptotically flat black
holes:
\citeNP{penrose-grav-coll-st-sings,hawking-ellis-lrg-scl-struc-st}),
or with trapping surfaces (for singularities contained in generalized
black holes: \citeNP{hayward-bh-confine,hayward-genl-laws-bhdyns}), or
with the ``boundaries'' of spacetime (such as big bang and big crunch
singularities), or they place the defining incomplete, inextendible
geodesic entirely in a compact subset of the spacetime
(\emph{e}.\emph{g}.,
\citeNP[pp.~290--292]{hawking-ellis-lrg-scl-struc-st}).  Singularity
theorems that are not pinpointing, such as those of
Gannon~\citeyear{gannon-sings-nonsimp-conn-sts,gannon-topo-slhs-sing-bhs},
merely demonstrate the existence of incomplete, inextendible geodesics
without giving one any information about ``where'' the geodesic is in
spacetime.

Now, the impact of possible violations will differ from theorem to
theorem depending on whether the theorem at issue pinpoints or not,
and on whether the violation can be localized in an appropriate sense
to that region of spacetime in which the theorem locates the predicted
singularity.  For theorems that do not pinpoint, I think there is no
principled reason to believe that any salient violations may or may
not vitiate the theorem.  For theorems that do pinpoint, there may be
hope of showing that at least some salient violations may or likely
will not vitiate the theorems, but one must work through them on a
case by case basis to make the determination.  If one has some reason
to believe, for example, that a given type of salient violation can be
segregated entirely from the region of spacetime in which a closed,
trapped surface forms and evolves (because, \emph{e}.\emph{g}., of the
type of collapsing matter that eventuates in the trapped surface),
then one also has some reason to believe that any theorem that both
invokes the violated condition and places the singularity in such a
closed, trapped surface may still hold despite the violation.  It
would take us too long to go through all the singularity theorems and
all the types of violations to determine which violations can and
cannot be relevantly segregated from the regions where the predicted
singularities form or reside.  I leave this as an exercise for the
reader.  

Similar considerations about pinpointing, type of violation, and the
possibility of segregation come into play when trying to determine
whether a given violation should give us reason to doubt the soundness
of any other type of given consequence of an energy condition.  I see
no way to draw clean, general conclusions.

In sum, it seems difficult to escape the conclusion that we are faced
with the horns of an important dilemma: either we must learn to live
with the ``exotic'' physics that violations of energy conditions lead
to (wormholes, closed timelike curves, sudden future singularities,
spatial topology change, naked singularities, \emph{et al}.), and so
become much more skeptical of the plethora of seemingly important
results that rely on the conditions; or else we must reconstruct
fundamental physical theory root and branch, \emph{e}.\emph{g}., by
prohibiting the use of essentially all scalar fields, in order to rule
out the possibility of such violations.  I personally find it more
realistic, if not more palatable, to grasp the first horn.  An
investigation of the consequences of this conclusion for projects that
purport to provide fundamantal explication and interpretation of the
conceptual and physical structure of general relativity is beyond the
scope of this paper, but is, I think, urgently called for.

\subsection{Appendix: The Principle of Equivalence}
\label{sec:princ-equiv}

There is an interesting, though not obvious, possible connection
between the principle of equivalence (in at least some of its guises)
and energy conditions.  Postulating the lack of a preferred flat
affine connection is, to my mind, one of the most promising ways of
trying to formulate the principle of equivalence in way that one can
make somewhat precise \cite{trautman65,trautman-gr}, even if one
cannot show that such a principle must be true in the context of the
theory.  Could one derive an energy condition, or the violation of
one, from the existence of a preferred flat affine structure?  One way
to determine such a privileged flat affine connection would be by use
of the existence of a distinguished family of particles possessing
what, for lack of a better term, I will call ``anti-inertial charge'',
which would couple with the ``active gravitational mass'' of ordinary
matter in such a way as to result in the anti-inertial systems
traversing curves whose images form the projective structure of a flat
affine connection.  For a force that picks out such a connection, one
can assign to it a stress-energy tensor by solving the equation of
geodesic deviation using it as a force that exactly cancels out the
curvature terms due to the ordinary affine connection, and deriving an
expression for an ``effective'' stress-energy tensor associated with
the force.    

One possible mechanism for producing anti-inertial charge is strongly
suggested by the arguments of \citeN{bondi-neg-mass-gr} showing that
active and passive gravitational mass is not necessarily equal in
general relativity, at least when negative mass is allowed.  In
particular, negative masses uniformly repel all other mass,
irrespective of the sign of the other masses, and likewise that
positive masses uniformly attract all other masses, and so, most
strikingly, a system consisting of one positive and one negative mass
will uniformly accelerate.  In this case, negative mass plays the role
of an anti-inertial charge.  Arguably, the inequality of passive and
active gravitational mass already constitutes a violation of the
principle of equivalence, at least in one of its guises.  In the case
that Bondi describes, therefore, the projective structure of the flat
affine connection could possibly be determined by the acceleration
curves of systems having equal parts positive and negative active
gravitational mass.  

This line of thought suggests the following.
\begin{conjecture}
  \label{conj:flat-affine-viol-ec}
  If one were able to demonstrate the existence of a privileged flat
  affine connection, by the existence of a family of particles with
  anti-inertial charge, then one or more of the standard pointilliste
  energy conditions would be generically violated.
\end{conjecture}

\subsection{Coda: The Trace Energy Condition}
\label{sec:coda-trace-econd}

\resetsub

The history of what may be called the Trace Energy Condition (TEC)
should give one pause before rejecting possible violations of the
standard energy conditions on the grounds that the circumstances or
types of matter involved in the violations seem to us today ``too
exotic''.  The TEC states that the trace of the stress-energy tensor
can never be negative ($T = T^n {}_n \geq 0$---or, depending on one's
metrical conventions, that it can never be positive).  In its
effective formulation, therefore, the condition requires that $p \leq
\athird \rho$ in a medium with isotropic pressure.  Before 1961, it
seemed to have been more or less universally believed in the general
relativity community that this condition would always be satisfied,
even under the most extreme physical conditions.  It is, for instance,
assumed without argument, or even remark, in the seminal papers of
\citeN{oppenheimer-volkoff-mass-neutron-cores} and
\citeN{harrison-et-report} on possible equations of state for neutron
stars.  It was not seriously questioned until the work of Zel'dovich
in the early 1960s, in which he showed that a natural solution for a
quantum field theory relevant to modeling the matter in neutron stars
leads to macroscopic equations of state of the form $p =
\rho$.\footnote{See \citeN{zeldovich-novikov-stars-rel} (especially
  p.~197) for a discussion.}  In fact, it is widely believed today
that matter at densities above 10 times that of atomic nuclei, as we
expect to find in the interior of neutron stars, behaves in exactly
that manner
\cite[ch.~8]{shapiro-teukolsky-bhs-wdwarfs-neut-stars}.\footnote{This
  coda was inspired by the discussion in
  \citeN{morris-thorne-worms-st-interstell-trvl}.}

\section{Temporal Reversibility}
\label{sec:temp-revers}

\resetsec

For the purposes of the discussion in
\S\ref{sec:constraints-char-st-theors}, and because it is of some
interest in its own right, I will briefly discuss the relation of the
energy conditions to the idea of temporal reversibility.

A spacetime is temporally orientable if one can consistently designate
one lobe of the null cone at every point as the ``future'' lobe.  A
temporal orientation then is logically equivalent to the existence of
a continuous timelike vector field $\xi^a$; by convention, the future
lobe of the null cone at each point is that into which $\xi^a$ points,
and a causal vector is itself future-directed if it points into or
lies tangent to the future lobe.  To reverse the temporal orientation
is to take $-\xi^a$ to point everywhere in the ``future'' direction.
If $T_{ab}$ is the stress-energy tensor in the original spacetime,
then we want the time-reversed spacetime to have the stress-energy
tensor $T'_{ab}$ such that: the four-momentum of any particle as
determined relative to any observer will be reversed in the
time-reversed case; and the energy density of any particle as
determined relative to any observer will stay the same.  Formally,
\begin{enumerate}
    \item $T'_{an} (-\xi^n) = - T_{an} \xi^n$
    \item $T'_{mn} (-\xi^m) (-\xi^n) = T_{mn} \xi^m \xi^n$
\end{enumerate}
Clearly, then, $T'_{ab} = T_{ab}$.  So, in sum, I claim the rule for
constructing the time-reverse of a (temporally orientable)
relativistic spacetime is to leave everything the same except for the
sense of parametrization of timelike (and null) curves, which should
be reversed.  (Note that no problem arises with parametrization of
spacelike curves: there is no natural or preferred sense for their
parametrization in the first place.)

This makes physical sense.  The best way to see this is to ask what
should happen to the metric under time-reversal.  I claim the answer
is: nothing at all.  The metric stays the same.  Temporal orientation
is not a metrical concept.  It is a concept at the level of
differential topology and conformal structure.  The temporal
orientation is determined by how one parametrizes temporal curves
(which in turn, of course, depends on whether one can do so in a way
that consistently singles out a choice of ``future lobe of null cone''
at every point of the manifold in the first place).  It also makes
geometrical sense.  If one fixes a $1 + 3$ tetrad $\{
\overset{\mu}{\xi}{}^a \}_{\mu \in \{0, \, 1, \, 2, \, 3\}}$ (not
necessarily orthonormal) such that the metric at a point can be
expressed as $\sum_\mu \alpha_\mu \overset{\mu}{\xi}{}_a
\overset{\mu}{\xi}{}_b$, for some real coefficients $\alpha_\mu$, then
reversing the sign of $\overset{0}{\xi}{}^a$ clearly does not change
the metric.\footnote{Another way to see this is to note that the only
  reasonable choice for ``changing the metric'' under time-reversal
  would be to multiply it by $-1$; that however, does not change the
  Einstein tensor, and so \emph{a fortiori} cannot change the
  stress-energy tensor.}  (One can always find such a tetrad at a
single point, though it may not be extendible to a tetrad-field with
the same property.)

It is a simple matter to verify that a spacetime satisfies any one of
the standard energy conditions listed in
\S\ref{sec:standard-energy-conds} if and only if the time-reverse of
the spacetime does as well.  (The same holds as well for all the more
recently proposed energy conditions discussed in
\S\ref{sec:very-recent-work}.)  On the face of it, this is somewhat
surprising.  A white hole, for instance, is the time-reverse of a
black hole, and surely that should violate some energy condition.  But
in fact, no, it shouldn't, as a perusal of the relevant Penrose
diagram will show: a white hole will violate an energy condition if
and only if its time-reversed black hole does so.

\section{Constraints on the Character of Spacetime Theories}
\label{sec:constraints-char-st-theors}

General relativity assumes the existence of a single object, the
stress-energy tensor $T_{ab}$, that encodes, for all fields of matter,
all properties relevant to determining the relationship of the matter
to the geometrical structure of spacetime.  This relationship is
governed by the Einstein field equation,
\[
G_{ab} = 8 \pi T_{ab}
\]
This equation, conjoined with the definition of a spacetime model
$(\mathcal{M}, \, g_{ab})$, constitutes the entirety of general
relativity as a formal theory.

As its name suggests, the stress-energy tensor encodes for matter only
information about what we normally think of as its energy, momentum
and stress content.  General relativity, then, assumes that what we
normally think of as stress-energy content completely determines the
relation of spacetime structure to matter---no other property of
matter ``couples'' with spacetime structure at all, except in so far
as it may have a part in determining the stress-energy of the matter.
It is exactly this feature of general relativity that affords the
energy conditions their power.  Nonetheless, we fully expect, or at
least fervently hope, that general relativity will one day give way to
a deeper theory of gravity, one that will attend to the presumably
quantum nature of phenomena in regions of extreme
curvature.\footnote{I will not discuss the relation of energy
  conditions to any programs in quantum gravity, as I do not feel any
  of them are mature enough as proposals for a physical theory to
  support serious analysis of this sort.  See \citeN{curiel-modesty}
  for why I hold this view.  See \citeN{wuthrich-raiders-lost-st} in
  this volume, among others, for arguments to the contrary.}  It thus
makes sense to explore alternative theories of spacetime even in the
strictly classical regime, if only to get ideas about how to try to
modify general relativity in the search for that deeper theory.
Surely not everything is up for grabs, though.  Even in the attempt to
formulate alternative theories in the spirit of free exploration, some
core structure or set of structures must be retained in order for the
explorations to take place in the province of ``spacetime theories''.
What is that core?  Is there a single one?

In particular, for our purposes, the most important question is: what
must be true about the relation of stress-energy to the local and
global structures of spacetime for one to be able to formulate energy
conditions and to use them to derive results?  What, we are thus led
to ask, must a spacetime theory itself be like in order for it to be
able to exploit the fact that deep and extensive features of global
structure depend only on purely qualitative properties of
stress-energy?  Any field equations it imposes must be ``loose''
enough to respect this fact.  In particular, no global feature of the
geometry, as constrained by a theory's field equations, should depend
on anything but purely qualitative properties of stress-energy;
\emph{a fortiori}, no global feature of the geometry should depend on
the species of matter present, so long as that species manifests a
relevant qualitative property.  It is otherwise difficult to see how
generic, purely qualitative conditions could determine specific,
concrete features of spacetime geometry.

A useful way to begin to try to address these questions, and at the
same time to begin to figure out the place of energy conditions in
relation to potentially viable alternative spacetime theories, is to
ask oneself, following a line of questioning introduced early in
\citeN{geroch-horowitz-glob-struc}, what one can envisage needing to
hold onto in future developments of physical theory, come what may.
Not the Einstein field-equation itself, most likely.  Very likely
causal structure of some sort.  What else?

What follows is my attempt at such a list of structures, roughly
ordered by ``fundamentality''---where I mean by this only something
like: what we would or should be willing to give up before what else,
what we have more and less confidence will survive in future theories
(not anything having to do with recent debates in the metaphysics
literature).  Such an ordering should respect, at a minimum, the fact
that one needs in place already some structure in order to be able to
define other structure---one could not countenance giving up the
former before the latter.\footnote{For a similar list, albeit
  constructed for a somewhat different purpose, and with a very
  different ordering than mine, see
  \citeN[p.~10]{isham-prima-fac-qs-qg}.}

In constructing the list, I have been guided by the tenet that any
physically reasonable spacetime theory should ``look enough like''
general relativity so as to make all the elements of the list make
sense in its context.  Not all the elements in the list, however,
should be understood to be restricted to the form they take in
standard accounts of general relativity.  For instance, ``causal
structure'' need not mean Lorentzian light cone structure; it may
signify, for example, only some relation among events required by some
feature of ambient matter fields, such as respecting the
characteristic cones of matter obeying symmetric, quasi-linear,
hyperbolic equations of motion, whether those cones conform to the
standard Lorentzian metric of spacetime.\footnote{See,
  \emph{e}.\emph{g}., \citeN{geroch-faster-light} and
  \citeN{earman-no-suplum-prop-class-q-flds}.}  Any such list,
moreover, will ineluctably be shaped in part by the biases, prejudices
and aesthetic and practical predilections of the one constructing it,
so the following attempt should be taken with a healthy dose of
salt.\footnote{One could sharpen this list by distinguishing between
  local and global varieties of structure, \emph{e}.\emph{g}., by
  allowing for the possibility that it makes sense to determine a
  local differential structure without necessarily requiring the
  existence of a global one.  While I think such distinctions could
  have interest for some projects, they are too \emph{recherch\'e} for
  my purposes here.}
\begin{enumerate}
    \item event structure: primitive set of ``events'' constituting
  the fundamental building blocks of spacetime\footnote{This does not
    presuppose that an ``event'' is a purely local entity, in any
    relevant sense of ``local''.}
    \item causal structure: primitive relation of ``causal
  connectability'' among events (not necessarily distinguishing
  between null and timelike connectability)
    \item topology: spacetime dimension; notion of continuous curves
  and fields (maps to and from event structure); relative notions of
  ``proximity'' among events; global notions of ``connectedness'' and
  ``hole-freeness'' on event structure
    \item projective structure, conformal structure, temporal
  orientability: notion of a set of events forming a ``straight
  line'', and so physically a distinguished family of curves (but not
  yet a distinction between accelerated and non-accelerated motion);
  distinction between null and timelike curves; preferred orientation
  for parametrization of causal curves; null geodesics (but not
  timelike or spacelike); asymptotic flatness; singularities
  (incomplete, inextendible causal curves); horizons (event, apparent,
  particle, \emph{etc}., and so asymptotically flat black holes)
    \item differential structure: notion of smooth (or at least
  finitely differentiable) curves and fields; and so of tangent
  vectors, tensors, Lie derivatives and exterior derivatives; and so
  of field equations and equations of motion; spinor structure
    \item affine structure: notion of accelerated versus
  non-accelerated motion, and so timelike geodesics; spacelike
  geodesics; ``hyperlocal'' conservation laws (covariant divergence),
  at least for quantities ``represented by'' contravariant indices on
  tensors; comparison (ratios) of lengths of curve-segments, and so
  integrals along curves
    \item metric structure: principled distinction between Ricci and
  Weyl curvature (``matter'' versus ``vacuum''); ``hyperlocal''
  conservation laws (covariant divergence) for any quantity; volume
  element, and so integrals, and so integral conservation laws (in the
  presence of symmetries) for spacetime regions of any dimension;
  variational principles; convergence and divergence of geodesic
  congruences (Rauchaudhuri equation), and so closed, trapped surfaces
  (generalized black holes)
    \item Einstein field equation: fixed relation between properties
  of ponderable matter and spacetime geometry; initial-value
  formulation and dynamics
\end{enumerate}

Now, granting the interest of the list for the sake of argument,
where, if at all, should one place energy conditions on it?  No matter
what else is the case, so long as definitional dependence (what one
needs in place already to define or characterize structure of a
particular sort) is one criterion used in ordering such a list, it
seems that energy conditions must be not so fundamental as
differential structure: one needs differential structure in order to
write down any tensor, and so \emph{a fortiori} to write down a
stress-energy tensor.  Because all the standard energy conditions (and
pretty much all the nonstandard ones), rely on the distinction between
causal and non-causal vectors in general, and often on the distinction
between null and timelike, it seems likely that energy conditions will
be less fundamental than conformal structure as well.  Energy
conditions, however, do not seem to require a notion of temporal
orientability, as the discussion of \S\ref{sec:temp-revers} strongly
suggests, and, except for the impressionist conditions, neither do
they require a projective structure.  They also seem to be largely
independent of topological structure (except in so far as it is
required to define differential structure).  The impressionist energy
conditions do require an affine structure (for the definition of a
line-integral along a geodesic), but since they have much murkier
physical significance and far fewer important applications than the
pointilliste ones, I would almost certainly prefer to forego them
before foregoing an affine structure.  

Now, if one accepts my ordering, or anything close to it, energy
conditions do not seem to fit anywhere neatly in it.  So what can we
conclude?  One possibility is that energy conditions are not clearly a
part of any broad conception of what a spacetime theory is, and thus,
perhaps, are not themselves of fundamental importance in the study of
the foundations of spacetime theories.  Alternatively, one could
choose to take the fact that energy conditions seem to fit nowhere
neatly in the list as a reason to change my groupings of structure
into levels or to change my proposed order of levels.

One reason to think they should form part of any broad conception of
what constitutes a spacetime theory rests on the remark of Geroch and
Horowitz I quoted on page~\pageref{pg:geroch-horo}, that without
energy conditions the Einstein field equation ``has no content.''  The
conditions one needs to impose to make the initial-value problem of
general relativity merely consistent---the so-called Gauss-Codazzi
constraints---look very much like conditions on the allowed forms of
types of matter.  So does the fact that the standard proofs showing
existence and uniqueness of solutions to the initial-value problem of
general relativity require matter-fields that yield quasi-linear,
hyperbolic equations of motion satisying something very much like the
DEC \cite{hawking-ellis-lrg-scl-struc-st,wald-gr}.  This fact seems to
place a constraint on spacetime theories---only theories that require
non-trivial input about the nature of matter in order for the
distribution of matter to constrain the geometry of spacetime ought to
be counted as physically reasonable, at least if we want to try to
hold on to the idea that a viable spacetime theory ought to support a
cogent notion of dynamical evolution, and thus (at a minimum) ought to
admit a well set initial-value formulation.

One can try to make this idea precise, and at the same time to capture
the kernel of Geroch and Horowitz's remark, in the following way.
First, note that globally hyperbolic spacetimes represent in a natural
way possible solutions to the initial-value problem of general
relativity as it is normally posed.\footnote{But see,
  \emph{e}.\emph{g}., \citeN{ringstrom-cauchy-prob-gr} for a
  discussion of the formidable subtleties and complexities involved in
  trying to make even this seemingly simple idea precise.}  Now, it is
a trivial matter to find globally hyperbolic spacetimes that violate
any energy condition.  Proof: pick your favorite globally hyperbolic
spacetime and some open set in it; from the formul{\ae} in
\citeN[Appendix~D]{wald-gr}, it follows that one can always find a
conformal transformation of the metric that is the identity outside
the open set and non-trivial inside such that at some point in the set
the transformed stress-energy tensor will yield whatever one wants on
contraction with a timelike or null vector; since conformal
transformations preserve causal structure, the transformed spacetime
is still globally hyperbolic.

Now, this fact poses a serious problem for any attempt to formulate a
notion of dynamical evolution that would support any minimal notion of
predictability or determinism.  Fix a Cauchy surface in the original
spacetime to the past of the open set one conformally jiggered in the
proof I sketched.  Take that Cauchy surface as initial data for the
initial-value problem of general relativity.  Which spacetime will the
Cauchy development off that Cauchy surface (the solution to the
initial-value problem with that initial data) yield?  The original
one?  One of the conformally jiggered one?  Another one entirely?  If
one cannot give principled reasons for why exactly one of those
spacetimes and no other is the natural result of dynamical evolution
off the Cauchy surface according to the Einstein field equation, then
one has captured one sense in which the Einstein field equation may
``have no content.''  The fact that the only known proof of the
theorem that a given globally hyperbolic extension of a spacetime is
the maximal such extension requires the WEC
\cite{ringstrom-cauchy-prob-gr}, in conjunction with the fact that
standard proofs of the well-posedness of the initial-value formulation
for general relativity rely on the DEC, suggest that it may be the
energy conditions that intervene to ensure a cogent notion of
dynamical evolution that supports some minimal notion of
predictability or determinism.

Holding on to everything in my list except for the Einstein field
equation, so long as whatever field equations do hold depend only on
something like the stress-energy tensor that does not depend on
idiosyncratic features of particular kinds of matter, I strongly
suspect that one will likely face the same problem.  Thus, once again,
we seem pushed toward the view that energy conditions play some
fundamental role or other in any reasonably broad conception of
spacetime theories, or at least any such conception that would include
a notion of dynamical evolution.

If one does think energy conditions belong as a part of any reasonably
broad conception of what constitutes spacetime theory, one tempting
way to try to capture the sense in which they may hold at a level of
structure deeper than the Einstein Field Equation invokes the
thermodynamical character of stress-energy: all stress-energy is
fungible, is inter-changeable, in the strong sense that the form it
takes (electromagnetic, visco\"elastic, thermal, \emph{etc}.), and so
\emph{a fortiori} any property or quality it may have idiosyncratic to
that form, is irrelevant to its gravitational effects, both locally
and globally.  This is not a conclusion that follows by logical
consequence from the observation that purely quantitative energy
conditions suffice to prove theorems of great depth and strength about
global structure.  It is only one that is strongly suggested by what
thermodynamics tells us about the nature of energy.  I will not be
able to discuss this idea further in this paper, however, as it would
take us too far afield.\footnote{In one of the first papers in which
  he tried to provide a fundamental derivation of the field equation
  bearing his name, \citeN[pp.~148--9]{einstein16/52} explicitly used
  a similar line of thought to motivate the idea that all
  gravitationally relevant mass-energetic quantities associated with
  matter of any kind is exhaustively captured by the stress-energy
  tensor:
  \begin{quote}
    The special theory of relativity has led to the conclusion that
    inert mass is nothing more or less than energy, which finds its
    complete mathematical expression in a symmetrical tensor of second
    rank, the energy-tensor.  Thus in the general theory of relativity
    we must introduce a corresponding energy tensor of matter
    ${T^\alpha}_\sigma$\ldots.  It must be admitted that this
    introduction of the energy-tensor of matter is not justified by
    the relativity postulate alone.  For this reason we have here
    deduced it from the requirement that the energy of the
    gravitational field shall act gravitatively in the same way as any
    other kind of energy.
  \end{quote}
}

The inability to derive the energy conditions from other propositions
of a fundamental character constitutes an essential part of what
pushes one to conceive of them as structure ``at a deeper level'' than
many other elements on the list, perhaps even deeper than causality
conditions (many of which can be derived from other fundamental
assumptions), and so applicable across a \emph{very} wide range of
possible theories of spacetime.  If, in the end, one does hold the
view that they ought to be thought of as a fundamental part of a
reasonably broad conception of what constitutes a spacetime theory,
then perhaps, as I suggested in \S\ref{sec:point-energy-cond}, the
final lesson here is that the geometric form of the energy conditions
are the ones to be thought of as fundamental, in so far as they rely
for their statement and interpretation only on invariant, geometrical
structures and concepts.  If that is so, then perhaps one potentially
fruitful way to use the (poorly named?) energy conditions as a
constraint on the construction of spacetime theories is to search for
theories in which these important geometric conditions have
unproblematic, physically significant interpretations.

\end{document}